\documentclass[10pt,journal,compsoc]{IEEEtran}
\ifCLASSOPTIONcompsoc
  \usepackage[nocompress]{cite}
\else
  \usepackage{cite}
\fi
\usepackage{url}

\ifCLASSINFOpdf
  \usepackage[pdftex]{graphicx}
\else
\fi
\usepackage{amsmath}
\newcommand{\calT}{\mathcal{T}}
\newcommand{\calI}{\mathcal{I}}

\newcommand{\calK}{\mathcal{K}}
\newcommand{\calL}{\mathcal{L}}
\newcommand{\calKL}{\mathcal{KL}}

\newcommand{\calV}{\mathcal{V}}

\usepackage{caption}
\usepackage{subcaption}
\usepackage{multirow}
\usepackage{amsfonts}
\usepackage{xcolor}

\newcommand{\blue}[1]{\textcolor{black}{#1}}

\begin{document}
%
\title{Exploring the Impacts of Power Grid Signals on Data Center Operations using a \\ Receding-Horizon Scheduling Model}
%
%
%
%

\author{Weiqi~Zhang, Line~A.~Roald,
        and~Victor~M.~Zavala
\IEEEcompsocitemizethanks{\IEEEcompsocthanksitem W. Zhang and V. M. Zavala are with the Department
of Chemical and Biological Engineering, University of Wisconsin-Madison, Madison,
WI, 53706.\protect\\
\IEEEcompsocthanksitem L. A. Roald is with the Department
of Electrical and Computer Engineering, University of Wisconsin-Madison, Madison,
WI, 53706.}
}

%
%

\markboth{Journal of \LaTeX\ Class Files,~Vol.~14, No.~8, August~2015}%
{Shell \MakeLowercase{\textit{et al.}}: Bare Demo of IEEEtran.cls for Computer Society Journals}
%



\IEEEtitleabstractindextext{%
\begin{abstract}
    Data centers (DCs) can help decarbonize the power grid by helping absorb renewable power (e.g., wind and solar) due to their ability to shift power loads across space and time. However, to harness such load-shifting flexibility, it is necessary to understand how grid signals (carbon signals and market price/load allocations) affect DC operations. An obstacle that arises here is the lack of computationally-tractable DC operation models that can capture objectives, constraints, and information flows that arise at the interface of DCs and the power grid. To address this gap, we present a receding-horizon resource management model (a mixed-integer programming model) that captures the resource management layer between the DC scheduler and the grid while accounting for logical constraints, different types of objectives, and forecasts of incoming job profiles and of available computing capacity. We use our model to conduct extensive case studies based on public data from Microsoft Azure and MISO. Our studies show that DCs can provide significant temporal load-shifting flexibility that results in reduced carbon emissions and peak demand charges. Models and case studies are shared as easy-to-use Julia code. 
\end{abstract}

\begin{IEEEkeywords}
Data centers, Load shifting flexibility, Scheduling, Rolling-horizon, Electricity markets, Mixed-Integer Programming 
\end{IEEEkeywords}}

\maketitle

\IEEEdisplaynontitleabstractindextext

%
\IEEEpeerreviewmaketitle

\IEEEraisesectionheading{\section{Introduction}\label{sec:introduction}}

\IEEEPARstart{T}{he} rapidly-growing computing and information industry is projected to consume 20.9\% of the total electricity by 2030 \cite{energy_use_growth}. This fast growth of electricity use (and associated carbon footprint) is being boosted by recent burgeoning of machine learning (ML) at rates as high as 10x per year \cite{amodei_2019}. A large fraction of these energy demands arise from hyperscale data centers (DCs), which are large computing facilities that are owned by IT companies and that are distributed across different geographical locations \cite{barroso_holzle_ranganathan_2018,dreyfuss_2018}. DC networks are managed in a coordinated manner, which allows computing tasks to be modulated/scheduled at multiple time scales and shifted geographically. In other words, the collective management of DCs allows for space-time shifting flexibility of computing, and hence power loads. This makes DCs a great source of flexibility that power grids and associated electricity markets can harness. However, current wholesale electricity markets are still designed around traditional large-scale generators and load utilities, which are not necessarily effective at incentivizing flexibility provision from DCs. On the other hand, big-tech companies are interested in direct participation in wholesale electricity markets and in reducing their carbon footprint. Compared to other common approaches to net-zero operations, such as local renewable generation, trading directly in wholesale markets allows for exposure to more renewable energy in a more affordable and less risky manner (a more through discussion on DC energy management options can be found in \cite{ghamkhari2016energy}). This has motivated Google, for instance, to participate directly in wholesale markets (in addition to using its 1.6 MW onsite solar generation facility at its  headquarters) \footnote{https://www.cnet.com/culture/google-gets-go-ahead-to-buy-sell-energy}.

To enable flexibility provision from DCs, it is critical to re-design current electricity markets; such restructuring must  allow for flexibility provision without dispatching a power allocation schedule that significantly degrades performance of DCs. In doing so, it is necessary to develop operational models for DCs that capture objectives, constraints, and operational logic of DCs and that captures how those are influenced by interactions with the power grid. Various energy-aware, online scheduling models have been proposed for DCs \cite{atiewi2016review}, driven by various metrics such as sustainability \cite{marahatta2020eco, paul2015demand}, fairness and latency \cite{polverini2013thermal, ren2012provably}. While these models and algorithms provide valuable insights into DC scheduling behavior, it is technically challenging to consider a setting where the scheduler interacts directly with electricity markets in real-life due to the high complexity of the DC scheduling logic. This becomes even more challenging if we consider high-intensity job profiles, as those found in cloud computing settings \cite{cortez2017resource}. As a result, a decision-making interface between DC schedulers and electricity markets is needed that captures scheduling flexibility in a simplified manner. The development of low-fidelity models to capture flexibility and interactions with the power grid has been explore in diverse application domains that include batteries \cite{sorourifar2018integrated, parvar2019analysis}, buildings \cite{ottesen2016prosumer}, electric vehicles \cite{zhang2020electric, jin2013optimizing, vagropoulos2013optimal}, and energy-intensive manufacturing processes \cite{otashu2018grid}. The computing industry is no exception; for instance, Google operates their DCs with a resource management layer between its scheduler and electricity markets \cite{radovanovic2021carbon}. Here, the scheduler is modeled directly as a combination of inflexible and flexible power loads and diverse details of the scheduling layer (e.g., computing delays) are ignored in order to achieve scalability. The resource management layer acts as an interface that sets power consumption targets for the scheduler based on various signals from electricity markets. 

\blue{In this work, we develop model for resource management of DC systems that enables  studies on load-shifting flexibility of DCs and of behavior arising at the interface between DCs and power grids. DCs benefit from being flexible in various ways; in this paper we focus on the benefits of reduced energy cost (in terms of reduction in peak demand charge) and reduced carbon footprint. We also discuss the level of flexibility from DCs as part of load shifting. We propose a receding-horizon optimization framework that captures power load allocations, costs, and emissions at the resource management level, while taking into account the discrete nature on the scheduling level. Specifically, the resource management level captured in our model decides the number of active servers available for the scheduling level. We emphasize that our framework is not meant to be an actual on-line scheduling algorithm for DCs (which should involve many other sources of complexity). Instead, our modeling framework is meant to be used as a high-level abstraction to study the effect of many external factors (e.g. electricity market signals) on DC behavior and to understand the degree of flexibility that DCs can offer to the power grid. We also highlight that, while the model assumes direct interaction with wholesale markets, it can readily be extended to study interaction with microgrids (e.g., that embed on-site renewable generation). Moreover, our model is computationally scalable, and can thus be embedded within power grid optimization models used for long-term planning and unit commitment.}

Our framework uses mixed-integer programming to model DC job scheduling and power allocation logic. The framework uses a state-space representation that captures the different states of computing jobs and DC servers, which is inspired by formulations for chemical process scheduling \cite{subramanian2012state}. Compared with Google's resource management model reported in \cite{radovanovic2021carbon}, our model considers the discrete nature of DC loads and considers uncertainty in an online setting. Rather than modeling each individual job, our framework embeds a flexible, aggregated representation for groups of jobs with the same size; this provides computational scalability to handle a large number of jobs (as seen in real trace data). At each time step, our model determines an optimal schedule for the next few hours; this receding-horizon framework allows us to capture realistic situations in which only limited/inaccurate forecast information is available on key parameters such as electricity prices, carbon emission rates, and incoming jobs. We use this framework to conduct extensive studies that aim to understand how much load-shifting flexibility DCs can provide and how power grid signals impact scheduling operations. Our results show that carbon-aware hyperscale DCs are able to offer significant flexibility to respond to carbon emission rate signals and peak demand charges, without significant performance loss (as guaranteed by job clearance constraints). In other words, the change in the operational trajectory leads to significant reduction in carbon emissions. We also find that the decision horizon at each time step is crucial to ensure stability and quality of short-term scheduling decisions; therefore, load forecast information is important and might affect the ability to participate in electricity markets. For capacity forecast errors, we find that job termination occurs when future capacity is overestimated. Finally, we find that DCs are able to respond to demand response signals that are revealed in a short time using a quadratic objective function.  All models and results are shared as Julia code in \url{https://github.com/zavalab/JuliaBox/tree/master/DataCenterScheduling}.

\section{Summary of Related Work}

A wide range of studies have explored the potential of harnessing load flexibility from DCs. A major line of research is to exploit temporal flexibility through demand response programs \cite{ghatikar_ganti_matson_piette_2012, wierman2014opportunities}. More recently, a strategy for providing regulation service reserves with quality-of-service guarantees is provided in \cite{zhang2019data}. Additionally, several studies also consider spatial flexibility enabled by geographical load shifting \cite{rao_liu_xie_liu_2010, rao_liu_ilic_liu_2012}. In particular, \cite{liu_lin_werman_low_andrew_2011, wang_ye_2016} explore how geographical load shifting facilitates renewable energy absorption, while \cite{kim2016data} find that a large faction of DC power loads can be fulfilled by using curtailed renewable power from the power grid (thus providing a mutual incentive for coordinating their operation). 

On the market side, several studies have explored whether electricity markets incentivize the provision of flexibility by DCs. The work in \cite{fridgen2017shifting} assesses economic incentives of geographical load-shifting across different electricity markets, while \cite{li2014modeling} and \cite{ruddy2014global} model how cost-sensitive DCs are incentivized by locational marginal price (LMP) signals. In a more general context, studies have explored general electricity market designs that exploit flexibility from demand and supply \cite{torbaghan2016local, ela2016wholesale}. A recent market design for remunerating load-shifting flexibility using virtual links is proposed in \cite{zhang2021remunerating}. In addition to LMP signals, some recent market designs propose to incentivize load shifting of DCs via carbon metrics \cite{lindberg2021guide, lindberg2020environmental}. Such market designs neglect operational constraints and logic of DCs that might prevent these assets from participating in the market (e.g., power allocations might lead to infeasible DC operation or high economic penalties). In addition, it is not clear whether DCs have an incentive to provide flexibility due to their internal interests of maximizing utilization to recover capital investments. 

Mathematical programming models are widely used in scheduling systems such as chemical manufacturing \cite{maravelias2012general, zhang2019review}, airline operations \cite{papadakos2009integrated}, and healthcare \cite{hall2012handbook}. These scheduling models produce an optimal schedule over a prediction horizon of interest, assuming that all information is known (e.g., demand forecasts). Recently, receding-horizon models have been developed to capture imperfect forecasts \cite{subramanian2012state}; in this approach, the scheduling system is interpreted as a model predictive controller (MPC). This MPC approach has been recently used to schedule energy-intensive systems such as buildings \cite{oldewurtel2012use} and batteries \cite{kumar2018stochastic}. MPC-like methods have also been applied to various DC scheduling problems, such as scheduling coupled with thermal management \cite{fang2014using, parolini2011cyber, fang2016qos}, risk-aware scheduling \cite{kusic2009power}, and set-point tracking \cite{wang2008cluster, xu2006predictive}. Our framework is different in that it acts as a resource management model that directly interacts with the power grid. Moreover, our framework allows for adaptive aggregation of job scheduling resolution (thus facilitating computational scalability). 

\section{DC Resource Management Model}

\begin{figure}
    \centering
    \includegraphics[width=0.4\textwidth]{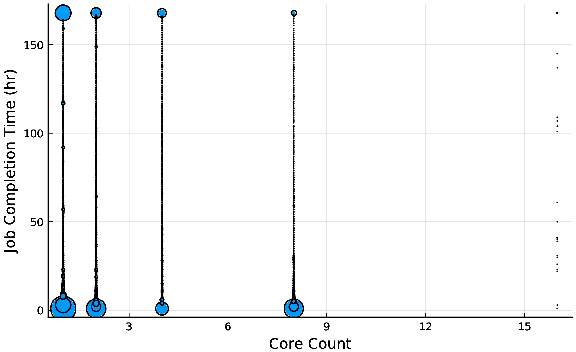}
    \caption{Grouping of jobs by resource requirement and completion time. Area of each circle is proportional to the number of jobs in the category each circle represents.}
    \label{fig:job_groups}
\end{figure}

We begin by considering the DC management problem with multiple computing servers $\calI = \{1, 2, ..., I\}$ running over a discrete time horizon $\calT = \{1, 2, 3, ...,T\}$. The DC manager does not necessarily have access to all $I$ servers at all times due to reasons such as limited power allocation. Thus, we define $I_t \in \mathbb{Z}_{\geq0}$ as the number of computing servers that are actually available at time $t$, with $I_t \leq I$. This parameter is dictated by the number of servers owned by the DC as well as by how much power is allocated by the electricity market. This is the key mechanism (for both offline and online scheduling) in which interaction with electricity markets come into effect. If the DCs receive more electric power from the market, they will have flexibility to turn on more servers (hence obtain higher $I_t$ values), and viceversa.

Each job is associated with a certain number of servers and a job runtime. To describe the incoming job profile, we let $\calK$ be the set of all possible numbers of requested servers, $\calL$ be the set of all possible job runtimes (in unit of hours), and $\calKL \subseteq \calK \times \calL$ be the set of all possible pair allocations. For example, figure \ref{fig:job_groups} shows an example of job trace data with $\calK = \{1,2,4,8,16\}$ and $\calL = \{1,2,3,...,168\}$. By grouping the computing jobs based on resource requirement and job runtime, we assume that all jobs are independent of each other so that the order of job execution does not matter. Under this assumption, the notion of "job" in this management model can also be used to capture a sequence of dependent computing tasks, which is aggregated and treated as one big job. This allows us to improve computational scalability by overlooking flexibility from scheduling-dependent jobs in different ways. In fact, we note that in our model we can tune the level of fineness/aggregation in the modeling (thus trading-off fidelity and scalability). This feature can dramatically simplify the problem and distinguishes our model from typical jobshop scheduling formulations. Instead of keeping track of the status of each single job at a time, our model considers the status of groups of jobs and treat all jobs in the same group homogeneously. 

We highlight here that the resource management layer makes decisions on the number of active servers available (denoted $m(t)$) for the scheduling layer based on power-related information from electricity markets. To enable such a decision-making mechanism, a relationship between server availability and power consumption is developed, and server availability is actually a proxy for power consumption. This means that the resource manager is deciding server availability and power consumption simultaneously. To simplify the complexity of solving the model, we assume a linear power consumption model where the power consumption is proportional to the amount of active servers: 
\begin{equation}
\label{eq:power_mapping}
    P(m(t)) = (P^{peak} - P^{idle}) \cdot \frac{m(t)}{I} + P^{idle}
\end{equation}
where $P^{peak}, P^{idle}$ are the peak and idle power of the DC. The energy consumption at each time can be computed as
\begin{equation}
\label{eq:energy_consumption}
    E(t) := P(m(t)) \Delta t
\end{equation}
where $\Delta t$ is the time interval (one hour in this paper). This adds another layer of versatility to our modeling framework, where the power mapping $E(\cdot)$ can be any model that can be embedded in an optimization problem. Common power models proposed in the literature include linear functions and piecewise linear functions \cite{radovanovic2021power}. From the power perspective, the load-shifting flexibility of DCs leads to flexibility of power dynamic trajectory for DCs. Delaying a computing job means reduction of the power consumption associated with executing that job at the current time and increase the power consumption at a later time. This is analogous to energy storage systems that schedule charge/discharge decisions. This idea can also be applied to geographical load shifting as a way to allocate power in space \cite{zhang2020flexibility}.

\subsection{Scheduling with Perfect Information and Fixed Power Availability}
\label{sec:perfect_info}

We first consider the scheduling problem with perfect information about incoming job profile and server availability. At the beginning of each time interval $t \in \calT$, a total number of $N_{kl}(t)$ jobs, that request $k$ computing servers and require $l$ hours to complete, are submitted to the DC. The scheduling decisions of the DC are $n_{kl}(t)$, which is the number of jobs initiated at the beginning of time interval $t$, and that require $k$ computing servers and $l$ hours to complete. The objective is to maximize the total number of active server-hours throughout the time horizon, which can be calculated from $n_{kl}(t)$ as:
\begin{equation}
    \sum_{t\in\calT} \left( \sum_{(k,l)\in\calKL} k \cdot l \cdot n_{kl}(t) \right)    
\end{equation}
This function aims to maximize server utilization by maximizing the number of active server-times. Equivalently, it is maximizing the so-called \textit{goodput} metric, which is the ratio of server-times for completed jobs to the total number of server-times available. In this model, loss in goodput incurs from unscheduled server-times only. The goodput metric is used to capture realistic behavior of the scheduling layer, which tends to maximize the utilization of computing resources to obtain maximum return on capital investment. In addition, the metric used here implicitly assumes that each job perfectly utilizes the server resource; as such, we are considering maximizing resource utilization at the job allocation level (not the utilization of a single server). 

The number of servers available at time $t$ is  denoted as$I_t$; this parameter is dictated by the number of servers owned by the DC as well as by how much power is allocated by the electricity market. This is the key mechanism (for both offline and online management) in which interaction with electricity markets come into effect. If the DCs receive more electric power from the market, they will have flexibility to turn on more servers (hence obtain higher $I_t$ values), and viceversa. The DC scheduling is subject to the following computing capacity constraint for each time $t\in \calT$:
\begin{equation}
\label{eq:capacity}
    \sum_{(k,l) \in\calKL} \sum_{t' = \underline{\tau}_{tl}}^t k \cdot n_{kl}(t') \leq I_t 
\end{equation}
where $\underline{\tau}_{tl} := \max\{t - l + 1, 1\}$. The left hand side calculates the number of active servers that are processing jobs at time $t$ as a function of $n$. By assuming perfect information, the DC has full knowledge of the values of $I_t$ throughout.

In order to ensure that a schedule obeys the submission time of jobs (i.e., a job cannot be started before it was submitted), we set up the following set of constraints for each time $t\in \calT$ and $(k,l) \in \calKL$:
\begin{equation}
\label{eq:job_submission}
    \sum_{t' = 1}^t n_{kl}(t') \leq \sum_{t' = 1}^t N_{kl}(t')
\end{equation}
This set of constraints ensures that, up to any time in the horizon, the schedule will not start more $(k,l)$ jobs than the number of jobs already submitted for each $(k,l) \in \calKL$. The above objective and sets of constraints lead to the following scheduling logic, defined over the  time horizon $\calT$:
\begin{subequations}
\label{opt:offline}
\begin{align}
    \max_{n} \quad & \sum_{t\in\calT} \left( \sum_{(k,l)\in\calKL} k \cdot l \cdot n_{kl}(t) \right) \\
    \text{s.t.} \quad & \sum_{(k,l)\in\calK\calL} \sum_{t' = \underline{\tau}_{tl}}^t k \cdot n_{kl}(t') \leq \bar{K}_t, \quad \forall \, t \in \calT \\
    & \sum_{t' = 1}^t n_{kl}(t') \leq \sum_{t' = 1}^t N_{kl}(t'), \quad \forall \, t \in \calT, \, (k,l) \in \calKL \\
    & n_{kl}(t) \in \mathbb{Z}_{\geq 0} \quad \forall \, t \in \calT, \, (k,l) \in \calKL
\end{align}
\end{subequations}
where $\mathbb{Z}_{\geq 0}$ denote the set of non-negative integers. The active server trajectory $m(t)$ can be computed as
\begin{equation}
    m(t) = \sum_{(k,l)\in\calKL} k \cdot n_{kl}(t)
\end{equation}

We want to note that our model can be seen as a generalization of Google's model in \cite{radovanovic2021carbon} in terms of load modeling. To one extreme, our model reduces to Google's model if no scheduling of discrete computing jobs is considered at all, which means that the power loads are assumed to be purely continuous. However, such continuous relaxation may overestimate the flexibility of DCs due to neglecting the discrete nature of the associated power loads.

\subsection{State-Space Representation of DC Management}
\label{sec:receding-horizon}

\begin{figure*}
  \centering
  \includegraphics[width=0.75\textwidth]{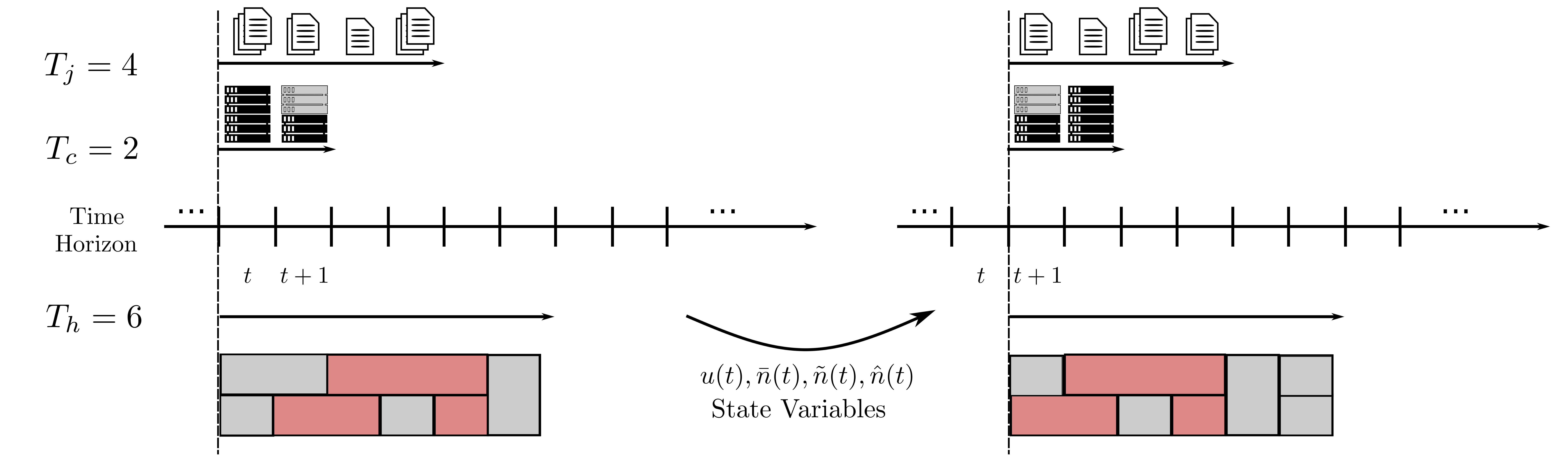}
  \caption{One step of the receding-horizon scheduling framework. Blocks at the bottom denote a feasible job schedule.}
  \label{fig:onestep}
\end{figure*}

Model \eqref{opt:offline} in Section \ref{sec:perfect_info} assumes that the resource management layer of DCs have perfect information over the time horizon about the computing availability $I_t$ and job submission profile $N$. In reality, however, accurate forecasts might only be available for a short period of time; for example, DCs can only typically predict incoming jobs and computing servers available in the next few hours. Moreover, the computing capacity $I_t$ is affected by the allocation and price of electric power delivered by the ISO. As a result of the uncertainty in the computing capacity, job termination might not be needed; under such scenario, running jobs have to be canceled if the real-time power allocated is not sufficient to run all computing servers needed. This type of correction/feedback/recourse logic behavior is not captured by model \eqref{opt:offline}.

The aforementioned uncertainty prompts DCs to refresh/correct job scheduling frequently. While a schedule is made at the beginning of the horizon, as time progresses the resource management layer observe what happens, obtain more accurate information about the future and need to adjust the schedule and, if necessary, terminate running jobs. To capture the this behavior, we propose a receding-horizon optimization framework for DC scheduling. Figure \ref{fig:onestep} shows different time scales of the receding-horizon framework. Here, $T_h$ denotes the decision horizon, in which a short-term schedule is decided at each step. Parameter $T_h$ captures how far ahead in time the optimization looks into the future; a shorter $T_h$ corresponds to a more greedy case. Parameter $T_j$ denotes the job forecast horizon, in which DCs are able to predict incoming job profile; $T_c$ denotes the capacity forecast horizon, in which DCs predict how many computing servers are available. The length of forecast horizons $T_j$ and $T_c$ capture one dimension of forecast ability; on the other hand, the accuracy of prediction can be arbitrary. We assume DCs always have perfect information about jobs and capacity at the current time interval; thus, $T_j = 1$ corresponds to the case of no job forecast (DCs only see and react as jobs arrive in the current interval), and $T_c = 1$ corresponds to the case of no server capacity forecast. We see that our framework can help capture a wide range of operational logic. 

\subsubsection{State Variables}
We now present a state-space representation of DCs. At each time $r \in \calT$, DCs determines a schedule for the shorter horizon $\calT_r = \{r, r+1, ..., r+T_h-1\}$. The model keep track of the following {\em state} variables for the DCs:
\begin{itemize}
    \item $u_l(r)$: number of servers that are committed for $l$ more timesteps at the beginning of interval $r$
    \item $\hat{n}_{kl}^{t_b}(r)$: number of $(k,l)$ jobs started at $t_b$ that are still running at the beginning of interval $r$. These jobs have run for $r-t_b$ time, and still need to run for $l - r + t_b$ times.
    \item $\Tilde{n}_{kl}(r)$: number of $(k,l)$ jobs waiting in queue at the beginning of interval $r$
    \item $\bar{n}_{kl}(r)$: number of completed $(k,l)$ jobs at the beginning of interval $r$
\end{itemize}
The state variables keep track of computing servers that are running (active), and the number of jobs that are waiting in queue, running, or completed (i.e., they keep track of ``inventories" of jobs and servers that are available and not available). The dimension indexed by the start time $t_b$ grows as the horizon progresses (as there are more possible start times to consider). The dimensions of the state variables are determined by the number of types of submitted jobs $|\calKL|$ and the current time $r$; precisely, by aggregating/disaggregating the jobs, we can manipulate the number of state variables that the resource manager keeps track of. The decision space of each state variable is $u(r) \in \mathbb{Z}^{\bar{L} - 1}_{\geq 0}$, $\hat{n}(r) \in \mathbb{Z}^{|\calKL| \times (r-1)}_{\geq 0}$, $\tilde{n}(r) \in \mathbb{Z}^{|\calKL|}_{\geq 0}$, $\bar{n}(r) \in \mathbb{Z}^{|\calKL|}_{\geq 0}$, where $\bar{L} := \max\{\calL\}$.

\subsubsection{Control Variables}
At each time $r$, the DC manager decides a schedule for the upcoming horizon $T_h$. Along with the jobs that are scheduled to start at time $r$, DCs also decide the number of running jobs to cancel at time $r$. This captures situations in which the operator has scheduled too many jobs in earlier time steps and the available computing servers are not sufficient to sustain the running jobs, which can happen if the available capacity is uncertain and capacity forecast is inaccurate. We refer to these variables as {\em control} variables, which are defined as follows.
\begin{itemize}
    \item $m(t)$: number of active servers at time $t \in \calT_r$
    \item $n_{kl}(t)$: number of $(k,l)$ jobs started at time $t \in \calT_r$
    \item $v_{kl}^{t_b}(r)$: number of $(k,l)$ jobs to cancel at the beginning of interval $r$ that begin at $t_b$ (so the leftover time is $t_b + l -t$ when canceled)
\end{itemize}
The dimension of the control variables are primarily determined by the number of types of submitted jobs ($|\calKL|$) and the current time $r$. In addition, the decision horizon $T_h$ also affects the size of the decision space. The decision space of each control variable is $m \in \mathbb{Z}_{\geq 0}^{T_h}$, $n \in \mathbb{Z}_{\geq 0}^{|\calKL| \times T_h}$, $v(r) \in \mathbb{Z}_{\geq 0}^{|\calKL| \times (r-1)}$. Figure \ref{fig:onestep} illustrates one step of our receding-horizon scheduling framework.

\subsection{Scheduling Logic}
We now detail the constraints and objective function of our scheduling model.

\subsubsection{Active Server Computation}
The number of active servers $m(t)$ at time $t$ can be computed as a function of job execution and termination decisions. Specifically, we include the following equality constraints for every $t \in \calT_r$:
\begin{multline}
\label{con:m_def}
    m(t) = \sum_{(k,l) \in \calKL} \sum_{t' = st(t,l)}^{t} k \cdot n_{kl}(t') + \sum_{l=t-r+1}^{\bar{L}-1} u_l(t) \\
    - \sum_{(t_b,k,l) \in \calV(t)} k \cdot v_{kl}^{t_b}(r)    
\end{multline}
where $st(t,l) = \max\{r, t-l+1\}$ denotes the earliest possible start time for a job of length $l$ to use computing resources at time $t$, and $\calV(t) := \{(t_b,k,l)\, | \, t_b +l-t > 0, (k,l)\in\calKL\}$ denotes the set of tuples $(t_b, k, l)$, such that the termination of $(k,l)$ jobs started at $t_b$ frees computing resources at time $t$. The first term computes the resource usage at time $t$ according to the new schedule. The second term accounts for the servers occupied by jobs scheduled in previous times. The third term computes the number of servers freed due to job termination. 

\subsubsection{Job Allocation}
At each stage, DCs can choose to schedule jobs already waiting in queue, or jobs newly submitted at the current time. This is captured by the following constraints.
\begin{equation}
\label{con:job_delay}
    \sum_{t'=r}^t n_{kl}(t') \leq \sum_{t'=r}^{t} N^e_{kl} (t') + \Tilde{n}_{kl}(r) \, \forall t \in \calT_r, (k,l) \in \calKL
\end{equation}
Here, $N^e$ denotes the general job forecast. In the case where data centers can have accurate jobs forecast within the forecast horizon, we have $N^e_{kl}(t) = N_{kl}(t)$ for $t < r + T_j$. Beyond the job forecast horizon $T_j$ no incoming jobs can be seen ($N^e_{kl}(t) = 0$ for $t \geq r + T_j$). In addition, any job forecast beyond the decision horizon does not affect the outcome of the current stage, as the schedule is only determined for $\calT_r$. We also note that constraints \eqref{con:job_delay} alone allow for unlimited temporal load shifting for jobs (i.e., they allow for an arbitrary job delay). 

\subsubsection{Minimum Job Clearance}
To ensure certain level of quality of service, we add constraints that specify a minimum number of jobs cleared (for each type of job).
\begin{equation}
\label{con:min_job_clear}
    \sum_{t'=r}^{r+T_h-1} n_{kl}(t') \geq \sum_{t'=r}^{r+\lfloor T_h / 2\rfloor-1} N^e_{kl} (t') + \Tilde{n}_{kl}(r) \quad \forall \, (k,l) \in \calKL
\end{equation}
Adding these constraints might lead to infeasible problems: if DCs do not allocate enough loads at early times, the queued jobs might make this constraint impossible to satisfy. Constraints \eqref{con:job_delay} and \eqref{con:min_job_clear} provide direct bounds on the operations of DCs. One can think of these constraints as upper bound and a lower bound trajectories for the number of active servers. Load-shifting flexibility is manifested as trajectories DCs can take within this flexibility envelope.

\subsubsection{Capacity Bound}
At each time interval, the number of active computing units is bounded by the computing capacity (i.e., the number of servers available $I_t$). At the current time $r$, the capacity information for future times may not be available; thus, we define $\bar{I}_t$ as the forecast for $I_t$. The capacity bound thus becomes
\begin{equation}
    m(t) \leq \bar{I}_t \quad \forall \, t \in \calT_r
\end{equation}
Note that $\bar{I}_r = I_r$ for each stage as the current information is assumed to be known accurately. In addition, if  DCs have accurate forecast for computing capacity at $t$, we have $\bar{I}_t$. 

\subsubsection{Job Termination Bound}
If DCs decide to terminate jobs, such actions are limited by the number of running jobs in the system at the current time $r$. These constraints are captured by the following inequalities:
\begin{equation}
    v_{kl}^{t_b}(r) \leq \hat{n}_{kl}^{t_b}(r) \quad \forall t \in \calT_r, (k,l) \in \calKL, t_b \in \calT^c_{rl}
\end{equation}
where $\calT^c_{rl} := \{\underline{\tau}_{rl},..., r-1\}$ denotes the possible start times for $(k,l)$ jobs that are still running at $r$. 

\subsubsection{State Variable Updates}
Up to now we have discussed operational constraints of online scheduling. The model needs to be updated at each time based on forecast and state information. We now introduce the state variable updates; at the end of each interval $r$, the value of each state variable at $r+1$ is updated based on the values of control variables $n(r), v(r)$ and the state variable value at $r$. The state variable $u_l$ are updated as follows for each possible length $l$:
\begin{equation}
\begin{aligned}
    u_l(r+1) = u_{l+1}(r) + \sum_{k|(k,l+1) \in \calKL} kn_{k,l+1}(r) \\
    - \sum_{t_b=1}^{r-1} \sum_{k | (k,l+r-t_b+1) \in \calKL } kv_{k,l+r-t_b+1}^{t_b}    
\end{aligned}
\end{equation}
The update rule propagates running jobs from the last time, accumulates newly submitted jobs and removes jobs of with remaining running time $l$ that are terminated at $r$. 

The number of $(k,l)$ running jobs is indexed by the start time $t_b$. First we note that jobs with $l = 1$ are not accounted for by the any state variable $\hat{n}_{kl}^{t_b}$ because they will finish running during the interval they are executed. Also, jobs started with $t_b \leq r - l + 1$ are not accounted for; there should be no such $(k,l)$ jobs running because they are either finished or terminated.

For the next stage, the number of running jobs with start time $t_b = r$ is accumulated based on the decisions $n_{kl}(r)$. Otherwise, the jobs will keep running at the next time unless they are terminated at $r$. Thus, the number of $(k,l)$ running jobs with begin time $t_b$ at $r+1$ is calculated as follows:
\begin{equation}
    \hat{n}_{kl}^{t_b}(r+1) = 
        \begin{cases}
            0 \quad \text{if} \quad l = 1 \text{ or } t_b \leq r - l + 1 \\
            n_{kl}(r)  \quad \text{if} \quad t_b = r \\
            \hat{n}_{kl}^{t_b}(r) - v_{kl}^{t_b}(r) \quad \text{o.w.}
        \end{cases}
\end{equation}

The number of $(k,l)$ jobs waiting in queue is updated as
\begin{equation}
    \Tilde{n}_{kl}(r+1) = \Tilde{n}_{kl}(r) + N_{kl}(r) - n_{kl}(r) + \sum_{t_b=r+1-l}^{r-1} v_{kl}^{t_b}(r)
\end{equation}
which accounts for new jobs that are not scheduled at $r$ and terminated jobs put back into the queue. The number of completed $(k,l)$ jobs follows a more complicated update rule: 
\vspace{-3pt}
\begin{equation}
    \bar{n}_{kl}(r+1) = 
        \begin{cases}
            \bar{n}_{kl}(r) + n_{kl}(r) \quad \text{ if } l = 1 \\
            \bar{n}_{kl}(r) + \hat{n}^{r-l+1}_{kl}(r) - v_{kl}^{r-l+1}(r) \text{ if } r-l+1 > 0 \\
            \bar{n}_{kl}(r) \quad \text{ o.w. }
        \end{cases} 
\end{equation}
If $l = 1$, jobs scheduled right at the beginning of interval $r$ will be finished in the end, and thus will be added to the completed jobs directly. For longer $l$, if the jobs begin at interval $r-l+1$, they will be finished at the end of interval $r$ unless they are terminated, thus the second case. In other cases, the number of completed jobs remains unchanged. 

\subsubsection{Objective Function}
The objective function for our model is flexible in that it captures multiple decision-making aspects of preference based on the desirable operation mode of DCs. At the baseline, DCs always seek to maximize the usage of their computing resources to clear all loads. This means that DCs prefer to clear as many jobs as possible, as early as possible, and avoid terminating jobs that are running (especially those that have been running for long). This motivates us to define the following function at each stage:
\begin{equation}
\begin{aligned}
    UTIL(r) := \sum_{t\in\calT_r}  \sum_{(k,l)\in\calKL} ((r+T_h) k l - t) \cdot n_{kl}(t) \\
    - \sum_{(k,l)\in\calKL}\sum_{t_b=\underline{\tau}_{rl}}^{r-1}((r+T_h) k l - t_b)\cdot v_{kl}^{t_b}(r)
\end{aligned}
\end{equation}
\blue{By maximizing this objective function, we find a set of control actions that maximize active server-hour (thus resource utilization) and minimize job termination within the decision horizon $\calT_r$. In addition, the objective function rewards executing jobs as early as possible, and puts more penalty on terminating jobs that have run for a longer time. This is achieved using the $(r+T_h)kl -t$ and $(r+T_h)kl - t_b$ weights for $n$ and $v$ variables, respectively. Specifically, $(r+T_h)kl -t$ assigns more reward for jobs scheduled earlier (with smaller $t$), and $(r+T_h)kl - t_b$ assigns more penalty for canceling jobs that have been running for longer (with smaller $t_b$). The $kl$ term in both weights serves to reward more for clearing jobs with higher server-hour values. The reward for early execution of jobs is necessary in order to ensure high resource utilization and avoid infeasibility that arises from incorporating minimum job clearance constraints \eqref{con:min_job_clear}, as mentioned earlier. This is especially true for cases in which the decision horizon is long and job forecast horizon is short, where the resource manager may be tempted to schedule early jobs to later times. Thus, the objective function can be thought of as posing an implicit penalty on job delay; on the other hand, the penalty on terminating long-running jobs prevent large goodput loss. }

In addition to resource utilization, DCs may be interested in being sensitive to other metrics. For instance, in order to be sustainable, DCs want to become aware of the carbon footprint as a function of their decisions at each stage; in order to be economically viable, DCs may want to minimize its energy use cost, which again is a function of their decisions at each stage. This will lead to multi-objective optimization formulations, which we convert to single-objective optimization problems using the weighted average of different constraints. 

We now discuss some specific metrics that can be included in the objective function; this is done in order to  highlight that our model is versatile. First, we need to extend the definition of the number of active servers $m(t)$ beyond $t \in \calT_r$ to future times to calculate the number of active servers based on the current schedule. This can be done in a similar fashion to defining $m(t)$ for $t \in \calT_r$, as in \eqref{con:m_def}. For convenience, we define $\bar{\calT}_r := \calT_r \cup \{r+T_h, ..., r+T_h+\bar{L}-2\}$ as the extended time horizon, which capture the effect of scheduling decisions $\calT_r$ in the near future.

We now extend the definition of $m(t)$ to $\bar{\calT}_r$. Specifically, for $t \geq r+T_h$, we add
\begin{multline}
\label{con:m_def_ext}
    m(t) = \sum_{(k,l) \in \calKL} \sum_{t' = st(t,l)}^{r+T_h-1} k \cdot n_{kl}(t') + \sum_{l=t-r+1}^{\bar{L}-1} u_l(t) \\
    - \sum_{(t_b,k,l) \in \calV(t)} k \cdot v_{kl}^{t_b}(r)    
\end{multline}
The only difference between \eqref{con:m_def_ext} and \eqref{con:m_def} is that, for \eqref{con:m_def_ext}, the second sum in the first term goes not to $t$ any more but to $r+T_h-1$ since there are no jobs scheduled beyond the horizon $\calT_r$. 

The reason we need to extend the definition of $m$ to future times is that we need these information to capture the effect of our short-term schedules on future times at the current time stage. For instance, scheduling long jobs that will, if not terminated, be running beyond the decision horizon length impacts the energy usage and carbon footprint in the near future.

We now discuss some of the specific metrics:
\begin{itemize}
    \item {\it Carbon Emissions.} The total carbon emission incurred by the scheduling decisions can be written as
    \begin{equation}
        \label{eq:ce}
        CE(r) = \sum_{t \in \bar{\calT}_r} cr(t) \cdot P(m(t)) 
    \end{equation}
    where $cr(t)$ is the carbon emission rate of the electricity generation fuel mix at time $t$. Note that by summing over the extended time horizon, we account for impacts on the times beyond the decision horizon.
    \item {\it Peak Demand Charge.} This captures a common type of charges for commercial customers, which depend on the maximum power within a period of time. For many commercial customers, peak demand charge can take 30 to 70 percent of their electric bills, in which cost is determined by the highest level of electricity used during a billing period, rather than the total amount of electricity consumption\footnote{https://www.cleanegroup.org/wp-content/uploads/Demand-Charge-Fact-Sheet.pdf}. To capture peak demand charges, we define $PD(r) = \max_{t \in \calT_r} P(m(t))$ as the peak demand for the current stage. This can be reformulated by adding the following constraints:
    \begin{equation}
        \label{eq:pd}
        PD(r) \geq P(m(t)) \quad \forall \, t \in \calT_r
    \end{equation}
\end{itemize}
We formulate the following objective function to capture all these metrics:
\begin{align}
    \label{eq:obj_fn}
    OBJ(r) &= UTIL(r) - \lambda_{CE} \cdot CE(r) - \lambda_{PD} \cdot PD(r)\nonumber
\end{align}
where $\lambda_{CE},\lambda_{PD} \in \mathbb{R}_+$ are the weighting parameters for each term. The choice of parameter values depends highly on specific case studies decision-making logic that one aims to capture. In the case studies that we explore later, we show how weights can influence behavior and we focus on emissions and peak demands. Other forms of costs, such as total costs rather than peak demand charge, can be easily captured by extending the objective function. We also note that, in this objective function, we are enforcing that $UTIL(r)$ is captured all the time. This is needed to ensure the logic of preferring scheduling jobs as early as possible. In summary, we formulate the following receding-horizon formulation at time interval $r$ with decision horizon $\calT_r$:

\begin{subequations}
\begin{align}
    \max_{m,n,u,v} \quad & OBJ(r) \\
    \text{s.t.} \quad & \eqref{con:m_def}-\eqref{eq:obj_fn}
\end{align}
\end{subequations}

\begin{figure}[h!]
     \centering
     \begin{subfigure}[b]{0.35\textwidth}
         \centering
         \includegraphics[width=\textwidth]{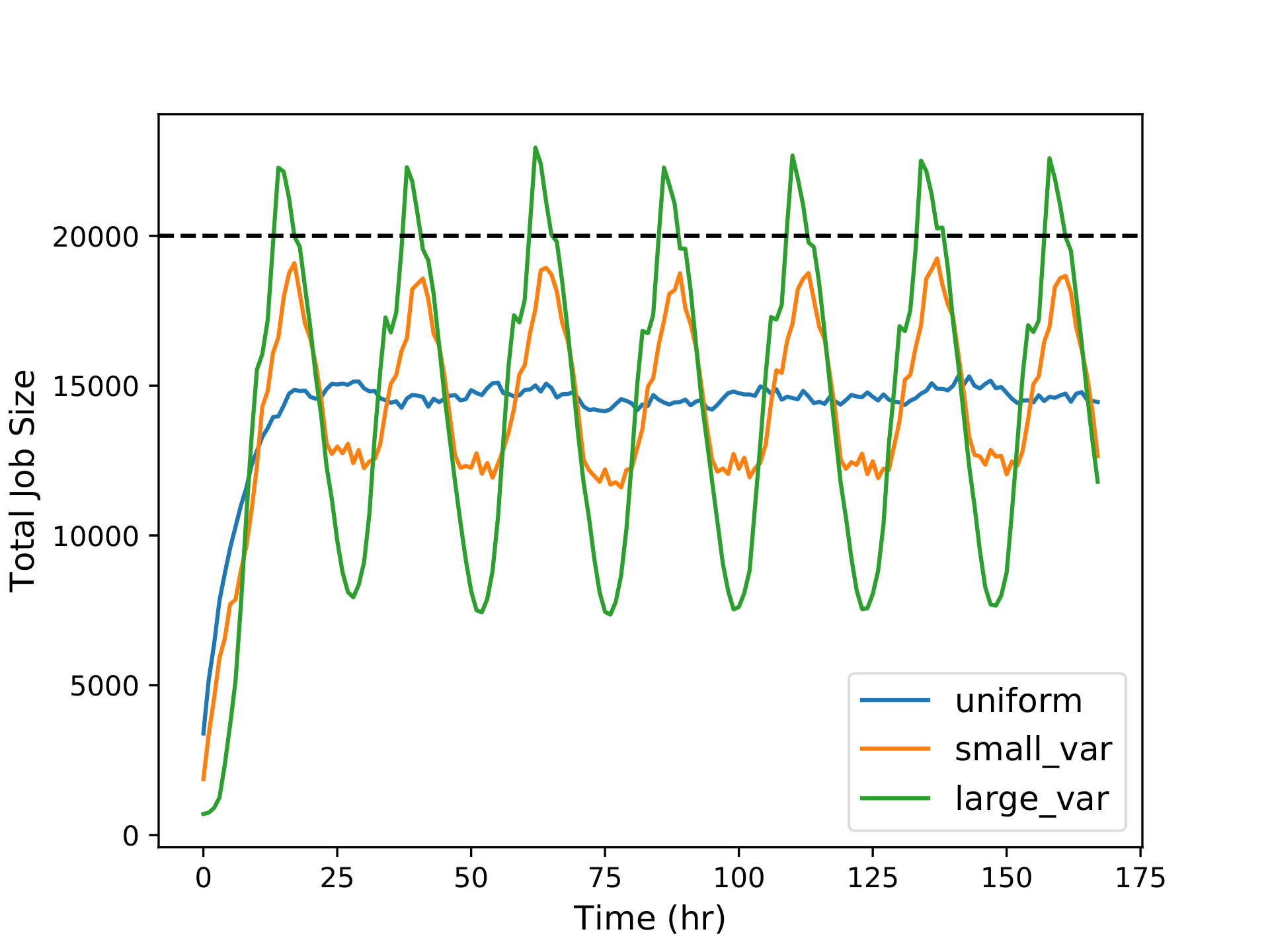}
         \caption{}
         \label{fig:job_profile}
     \end{subfigure}
     \hfill
     \begin{subfigure}[b]{0.35\textwidth}
         \centering
         \includegraphics[width=\textwidth]{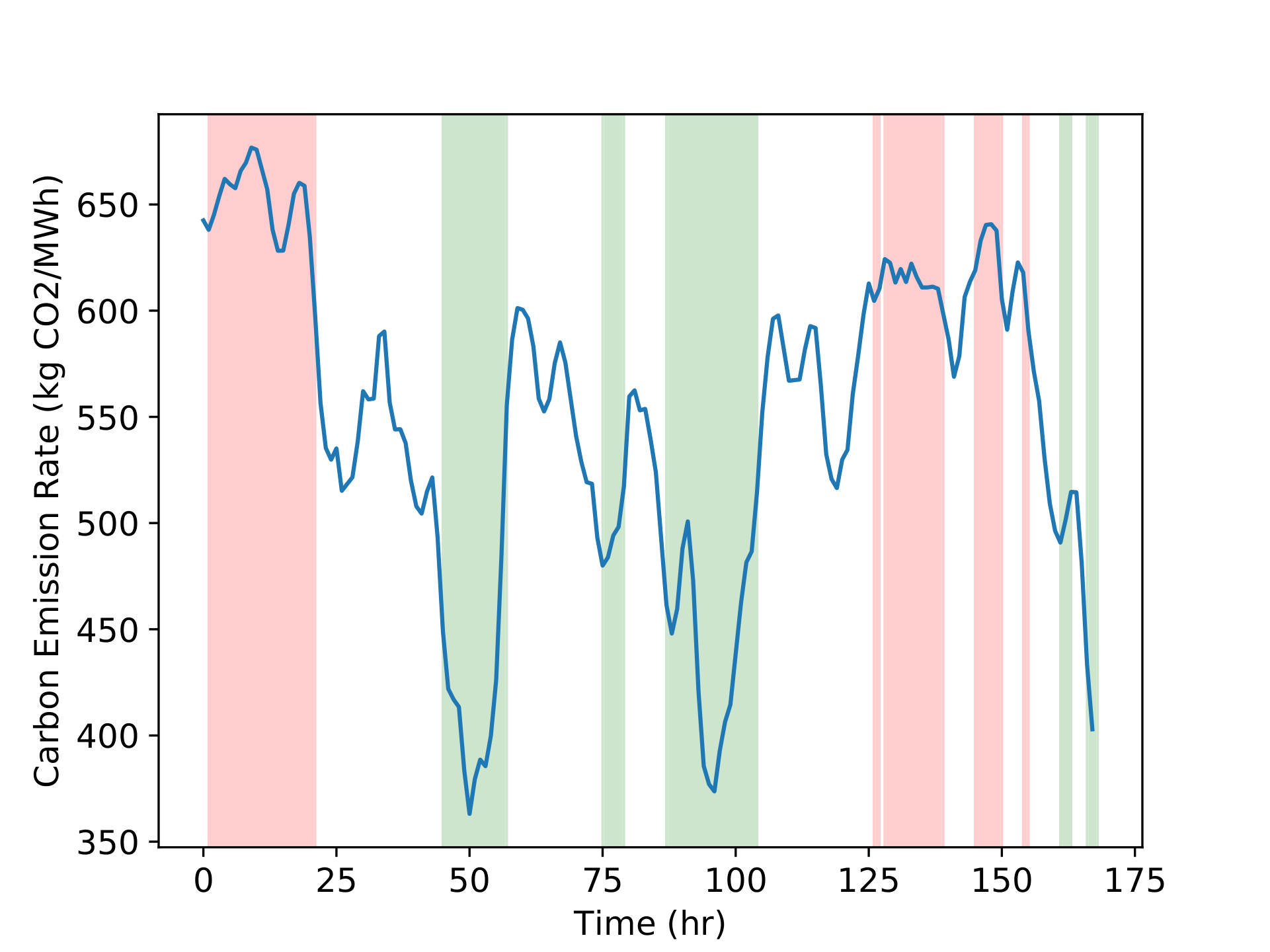}
         \caption{}
         \label{fig:cer}
     \end{subfigure}
    \caption{Job profile (a) and carbon emission rate (b) data}
    \label{fig:data}
\end{figure}

\begin{figure}[!t]
    \centering
    \includegraphics[width=0.35\textwidth]{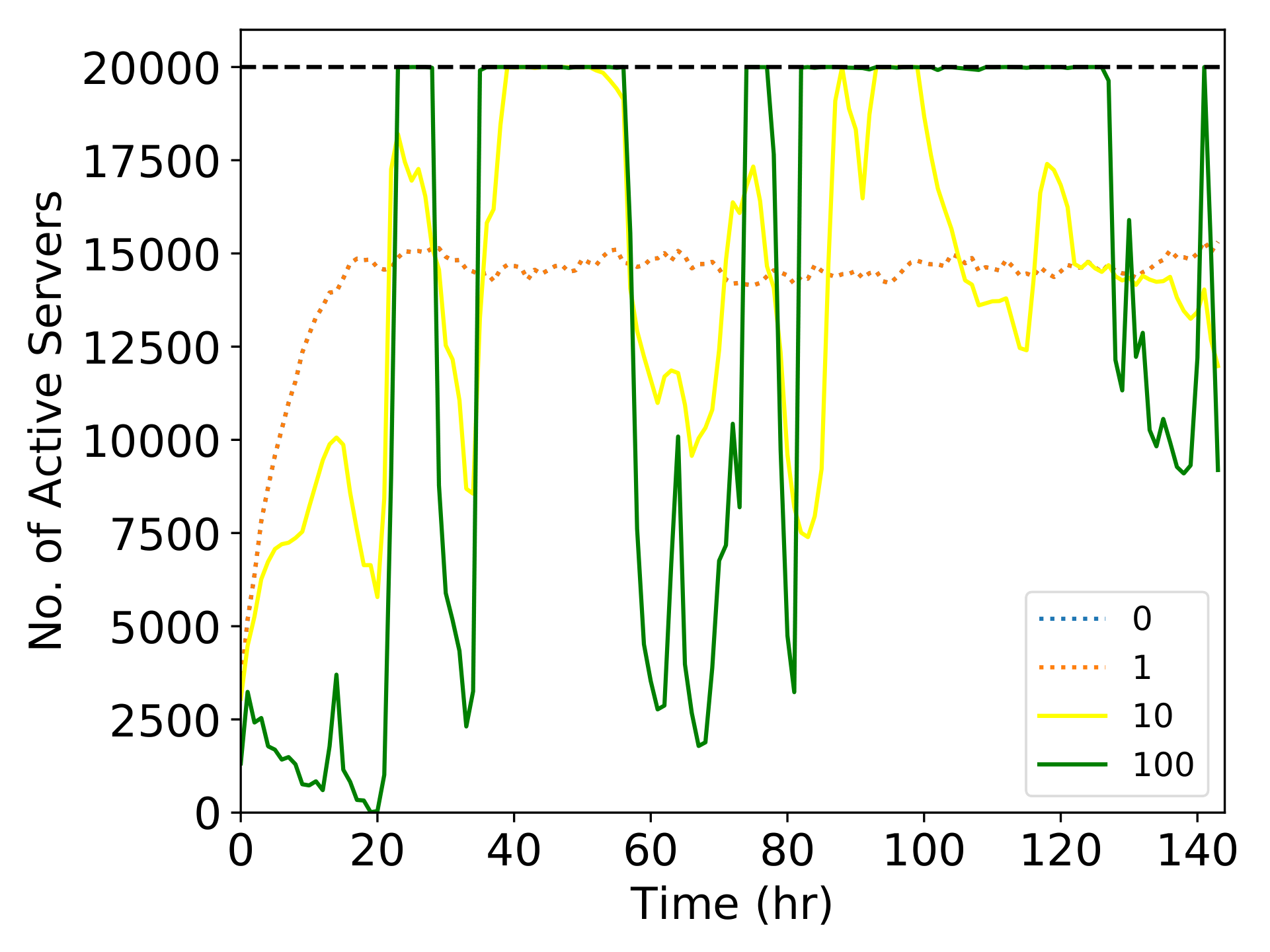}
    \caption{Active server trajectory with carbon awareness (legend denotes $\lambda_{CE}$ values)}
    \label{fig:ce}
\end{figure}

\begin{figure}[!t]
    \centering
    \includegraphics[width=0.35\textwidth]{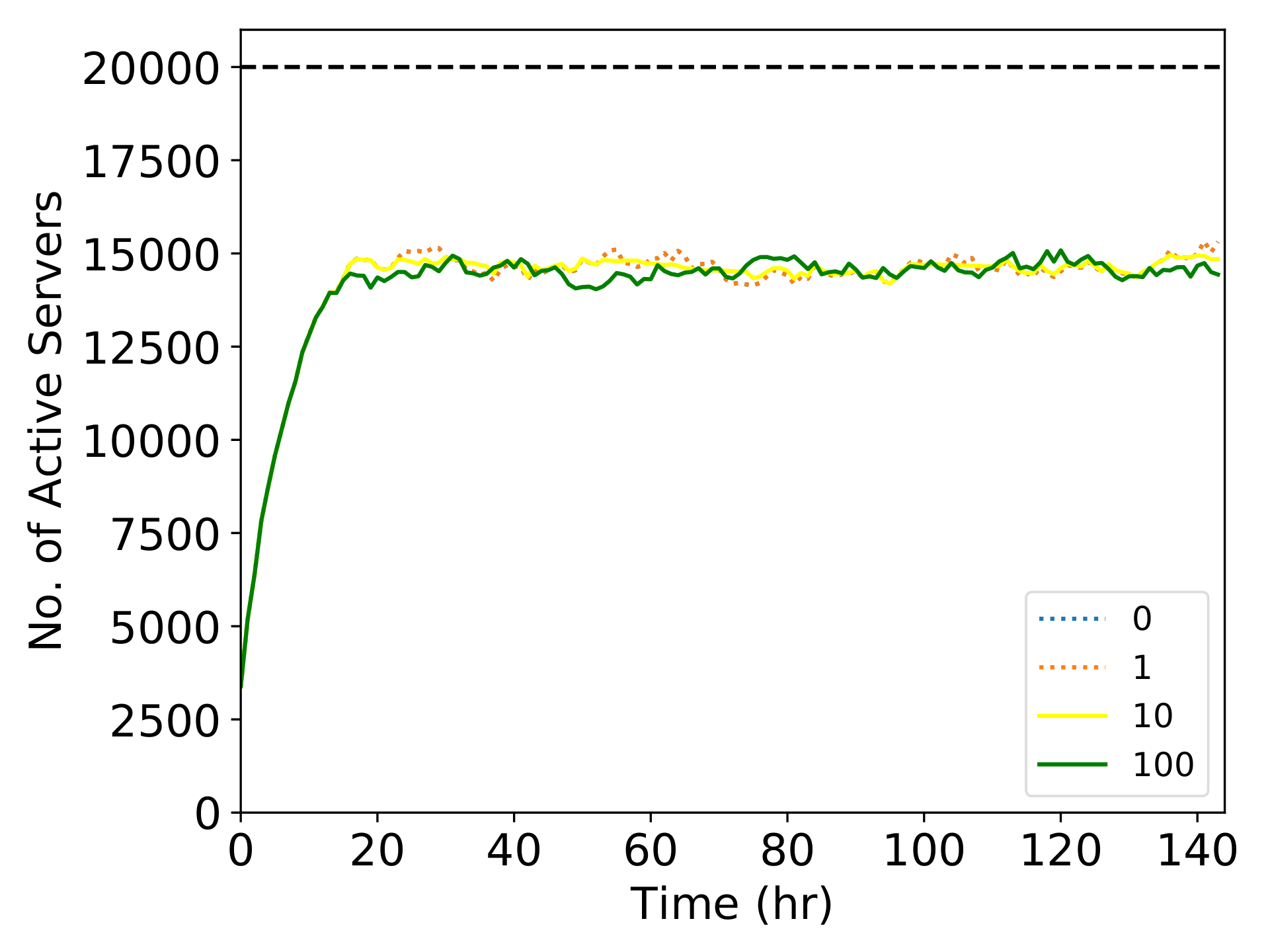}
    \caption{Active server trajectory with peak demand charge (legend denotes $\lambda_{PD}$ values)}
    \label{fig:pd}
\end{figure}

\section{Case Studies}

In this section we showcase the applicability of our model using case studies that help explore the temporal load-shifting flexibility of DCs. Specifically, we demonstrate benefits of shifting DC loads in terms of reduced peak demand charges and carbon emissions, which help DCs to reduce their electric bills and carbon footprint. We also discuss the level of flexibility demonstrated by DCs as part of load shifting, which we characterize as the extent of deviation from the baseline schedule. For our case studies, we consider a large DC system with 20,000 servers. The DC has a peak power of $P^{peak} = 100MW$ when all servers are running and idle power of $P^{idle} = 30MW$. For all our case studies, we run the model for 168 hours. 

We construct the incoming job data using the Microsoft Azure public dataset v1\footnote{https://github.com/Azure/AzurePublicDataset}. The dataset contains virtual machine trace data for a 30-day period in 2017. For our case studies, we filter out jobs running longer than 24 hours; implicitly we assume these are inflexible (delay-sensitive) jobs. We then sample job arrival time for each job from a probability distribution. We consider three types of incoming job profiles: uniform, small variance (\textit{small\_var}) and large variance (\textit{large\_var}). For the case of uniform, job arrival times are sampled from a uniform distribution; for the cases of small variance and large variance, job arrival times are sampled from a probability distribution with daily patterns. The corresponding computing load profiles are shown in figure \ref{fig:job_profile}. The electricity cost and carbon emission data are obtained from MISO public dataset from 07/25/2021 to 08/01/2021, as shown in Figure \ref{fig:cer}. Based on the considered level of job aggregation, each hourly step of optimization takes a few seconds to solve using Gurobi 9 on a cluster with Intel Xeon CPU E5-2698 v3. This highlights that our model can be used to capture scheduling logic while keeping computational times moderate.

\subsection{Response to Carbon Emission Rates and Peak Demand Charges}

\begin{figure*}[h!]
     \centering
     \begin{subfigure}[b]{0.33\textwidth}
         \centering
         \includegraphics[width=\textwidth]{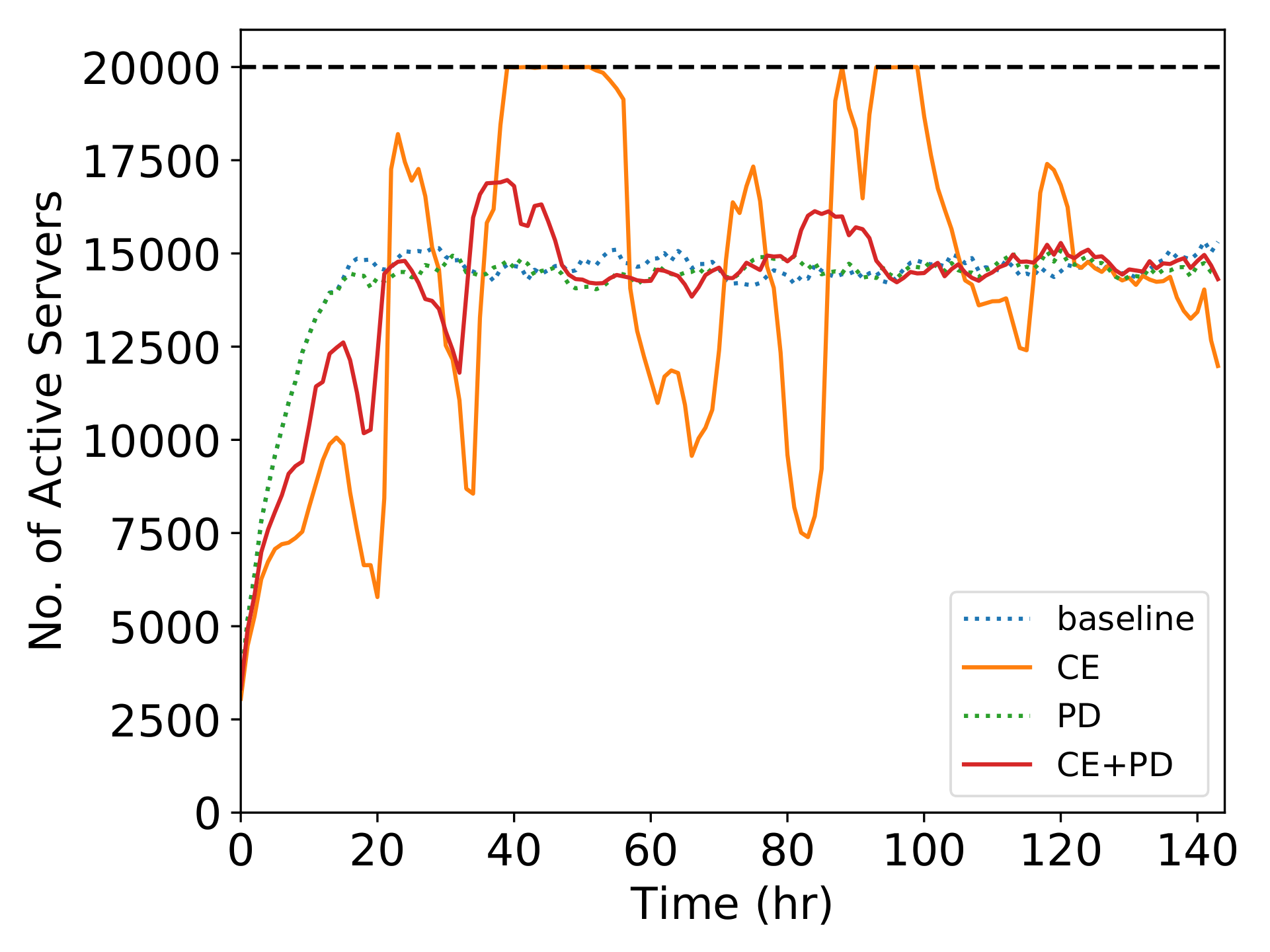}
         \caption{Uniform}
     \end{subfigure}
     \hfill
     \begin{subfigure}[b]{0.33\textwidth}
         \centering
         \includegraphics[width=\textwidth]{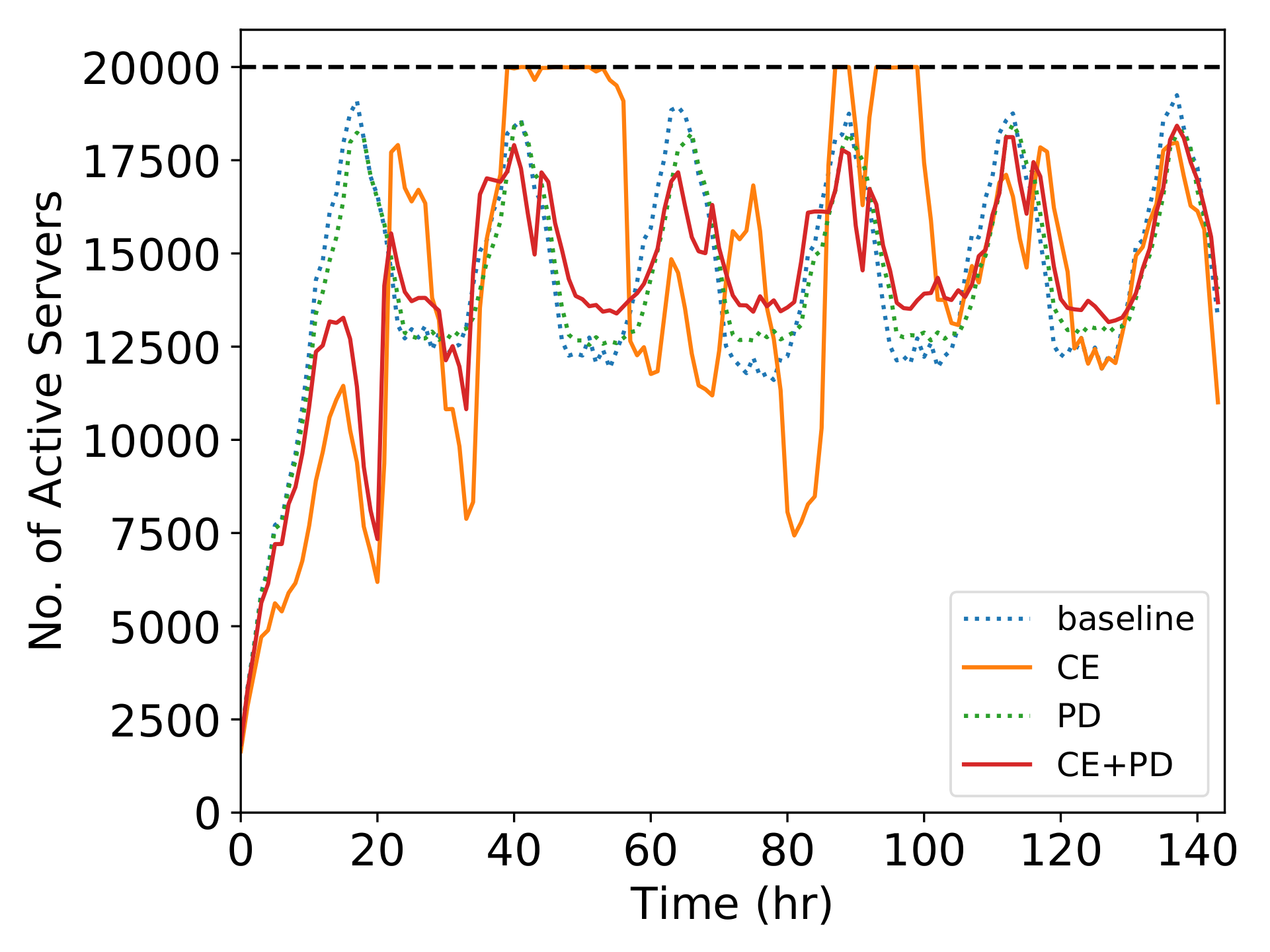}
         \caption{Small Load Variance}
     \end{subfigure}
     \hfill
     \begin{subfigure}[b]{0.33\textwidth}
         \centering
         \includegraphics[width=\textwidth]{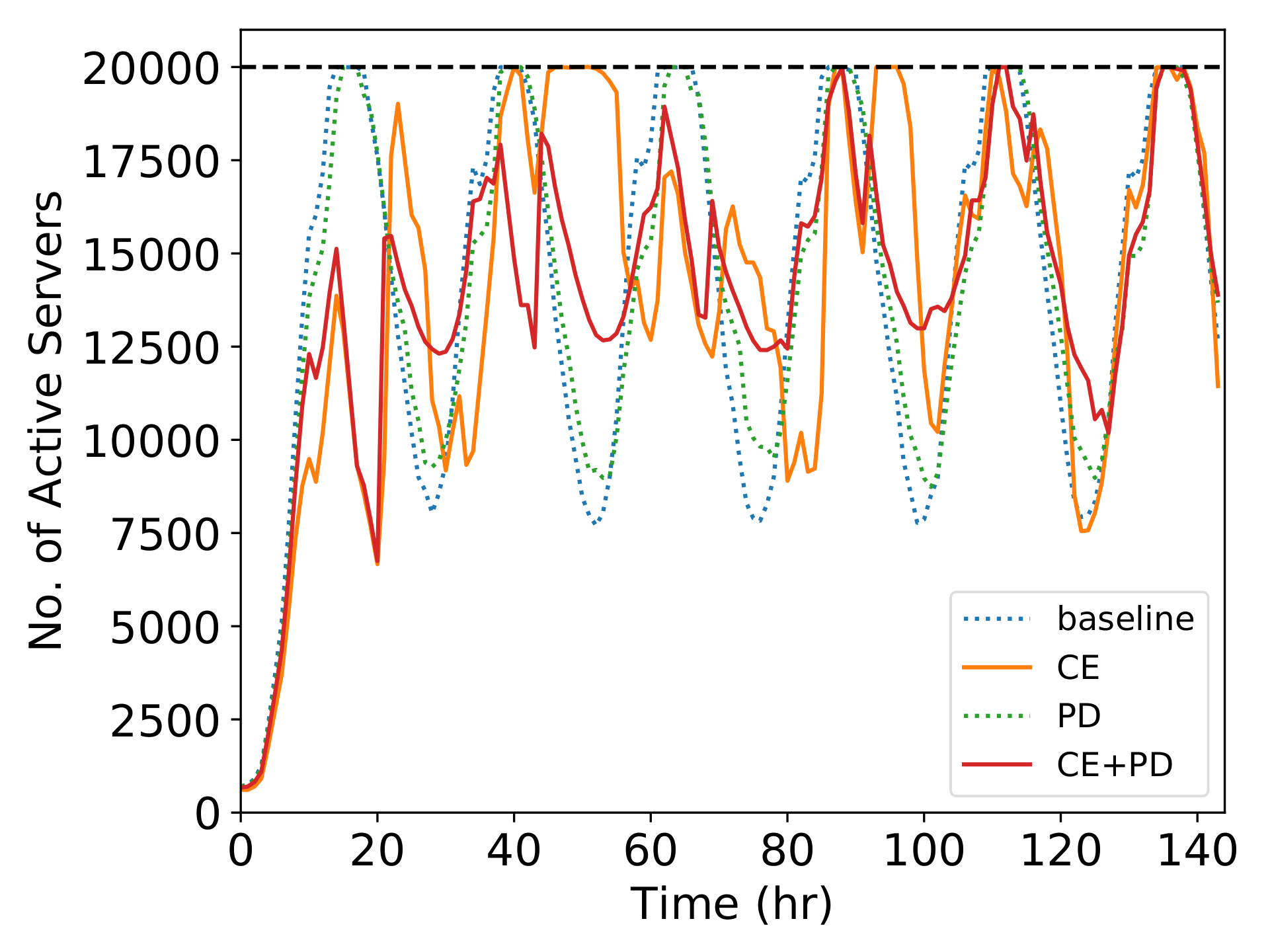}
         \caption{Large Load Variance}
     \end{subfigure}
    \caption{Active server trajectory with both peak demand charge and carbon awareness. Here, $\lambda_{CE} = 10$ for CE cases and $\lambda_{PD} = 100$ for PD cases, otherwise $0$. }
    \label{fig:pdce}
\end{figure*}

We explore how DCs adjust their operational trajectories in response to peak demand charge and carbon emission rate signals. We test different values of $\lambda_{CE}, \lambda_{PD}$. The decision horizon is fixed as $T_h = T_c = T_j = 24$, with perfectly accurate forecasts. 

We start by exploring different levels of sensitivity to carbon emission rates, as indicated by different values of the weight $\lambda_{CE}$. Figure \ref{fig:ce} shows the active server trajectories on the uniform job profile, where each case fixes $\lambda_{PD} = 0$ and takes a different value for $\lambda_{CE}$. We observe that the trajectory starts to exhibit significant deviation from the baseline ($\lambda = 0$) when $\lambda_{CE} = 10$. A trajectory with significant deviation from the baseline demonstrates great flexibility of the system, which is a result of significant load shifting in time. When DCs over-react to carbon emission signals ($\lambda_{CE} = 100$), the trajectory shows sharp ramping frequently. This shows the trade-off between stability and sustainability in DC operations (as frequent rapid ramping is generally not desired). We observe that, as $\lambda_{CE}$ is increased, more servers become active at times with low carbon emission rates and less active servers when high carbon emission rates. The fast ramping behavior at high $\lambda_{CE}$ values arises from volatility in carbon emission rates. This set of simple experiments shows that DCs are able to offer great flexibility if operated with an extremely environment-driven objective function.

We next explore different levels of sensitivity to peak demand, as indicated by different values of the weight $\lambda_{PD}$. Figure \ref{fig:pd} shows the active server trajectory on uniform job profile, where each case fixes $\lambda_{CE} = 0$. We observe that in all cases the trajectory is close to the baseline; even for large $\lambda_{PD}$ values, slight peak-shaving and trough-filling behavior is observed from the active server trajectories. This is expected as the baseline trajectory does not exhibit much variation. 

\begin{table}[h!]
\renewcommand{\arraystretch}{1.0}
\caption{Carbon emission and volatility results}
\label{table:pdce}
\centering
\begin{tabular}{|c|c|c|c|c|}
\hline
Job profile & $\lambda_{CE}$ & $\lambda_{PD}$ & $CO_2$ ($10^6$ kg) & $\sigma(m(t))$ ($10^3$)  \\
\hline
uniform & 0 & 0 & 6.25 & 1.72 \\
uniform & 10 & 0 & 6.11 & 4.31 \\
uniform & 0 & 100 & 6.23 & 1.70 \\
uniform & 10 & 100 & 6.19 & 2.24 \\
\hline
small\_var & 0 & 0 & 6.29 & 3.18 \\
small\_var & 10 & 0 & 6.13 & 4.48 \\
small\_var & 0 & 100 & 6.25 & 2.92 \\
small\_var & 10 & 100 & 6.20 & 2.90 \\
\hline
large\_var & 0 & 0 & 6.32 & 5.11 \\
large\_var & 10 & 0 & 6.17 & 4.87 \\
large\_var & 0 & 100 & 6.27 & 4.65 \\
large\_var & 10 & 100 & 6.21 & 3.93 \\
\hline
\end{tabular}
\end{table}

Figure \ref{fig:pdce} shows the active server trajectory with awareness of both peak demand and carbon emission rates, on all job profiles. Here, we assign $\lambda_{CE} = 10$ for CE cases and $\lambda_{PD} = 100$ for PD cases. For each case, we observe that the ramping becomes less frequent and less drastic with the addition of peak demand charge in the objective. The amount of carbon emission reduction for each case and the standard deviation for the active server trajectory as a measure of volatility is shown in table \ref{table:pdce}. The standard deviation is computed as $\sigma(m(t)) := \sqrt{\frac{\sum_{t=1}^{144}(m(t) - \bar{m})^2}{144}}$, where $\bar{m}$ is the average number of active servers. \blue{Expectedly, for each job profile, the case CE achieves the most carbon as it is tracking the carbon emission rates the most aggressively, as a result of significant load shifting driven by the carbon-aware objective function. Compared with CE cases, CE+PD cases attains lower maximum number of active servers, meaning that peak demand charge is reduced. Interestingly, the case CE+PD still achieves much carbon emission reduction compared with the baseline case, with much lower volatility $\sigma$. This shows that carbon emission rates, coupled with a peak demand charge structure, provide a reasonable incentive structure for DCs interested in achieving sustainable operations.}

\begin{figure}[!t]
    \centering
    \includegraphics[width=0.4\textwidth]{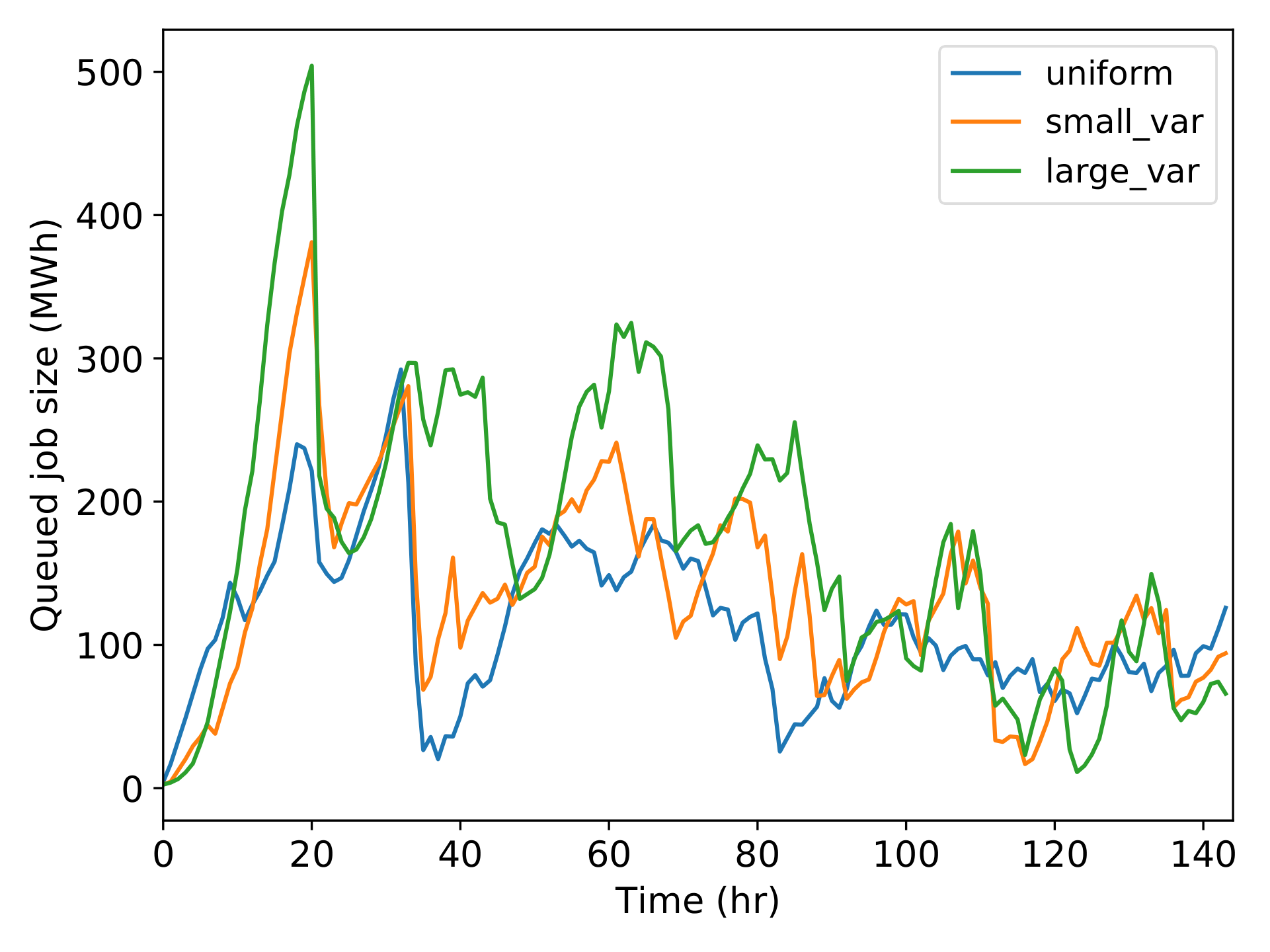}
    \caption{Total energy of queued jobs over time.}
    \label{fig:queued_job_size}
\end{figure}

\begin{figure}[!t]
    \centering
    \includegraphics[width=0.39\textwidth]{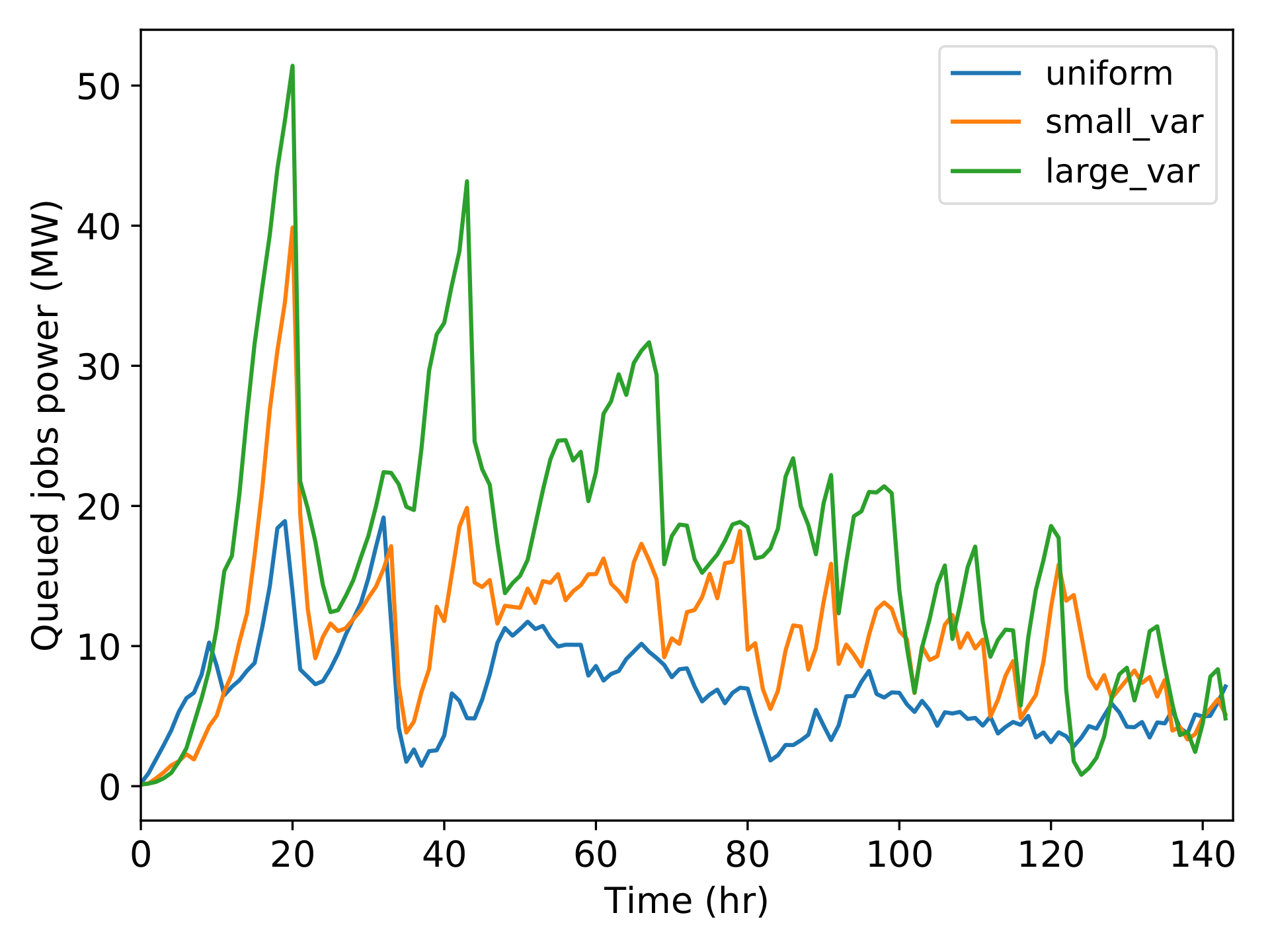}
    \caption{Total power of queued jobs over time.}
    \label{fig:queued_job_power}
\end{figure}


Figures \ref{fig:queued_job_size} and \ref{fig:queued_job_power} show how the energy needed to clear of job queue of our DC system changes over time for the case where both carbon emission and peak demand charge are considered. The energy requirement of queued jobs is measured as the total energy consumption of all queued jobs, which depends on the linear power mapping \eqref{eq:power_mapping} in our model. Similarly, the power of queued jobs is measured as the total power consumption of all queued jobs, which is a function of servers requested. Queued jobs of DCs are analogous to the state-of-charge of a battery system, except that jobs represent electricity consumption while state of charge characterizes electricity supply. We observe that, in all cases, the DCs experience an accumulation phase in roughly the first 20 hours, so as to store enough jobs for ramping up electricity load later. After the initial accumulation phase, DCs enter an oscillation mode, where the electricity load of DCs ramps up or down at different times. Furthermore, with more volatility from incoming jobs, queued jobs ramp in a wider range with a higher power of queued jobs. 

\subsection{Effect of Carbon Signal Accuracy}

\begin{figure*}[h!]
     \centering
     \begin{subfigure}[b]{0.33\textwidth}
         \centering
         \includegraphics[width=\textwidth]{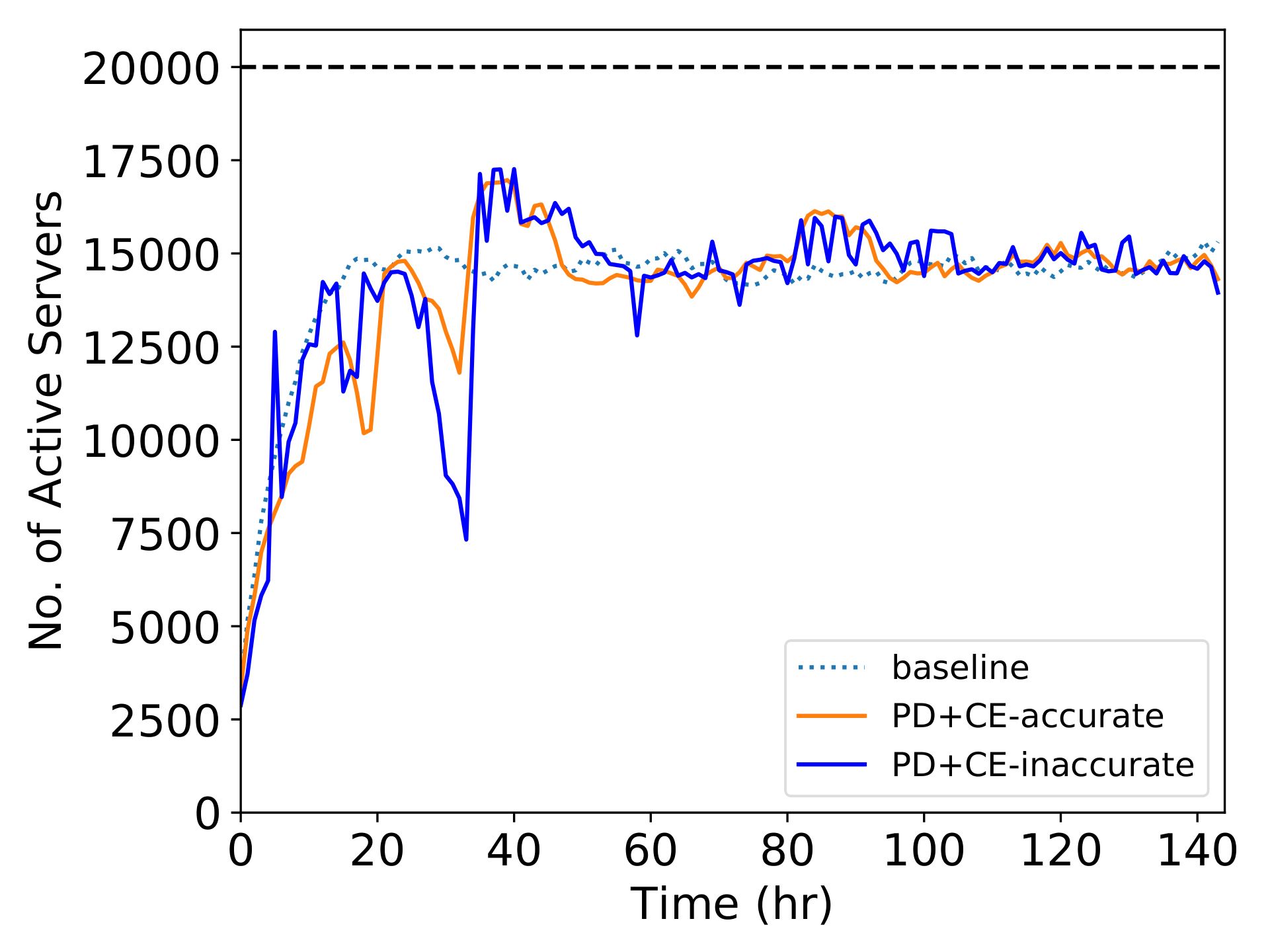}
         \caption{Uniform}
     \end{subfigure}
     \hfill
     \begin{subfigure}[b]{0.33\textwidth}
         \centering
         \includegraphics[width=\textwidth]{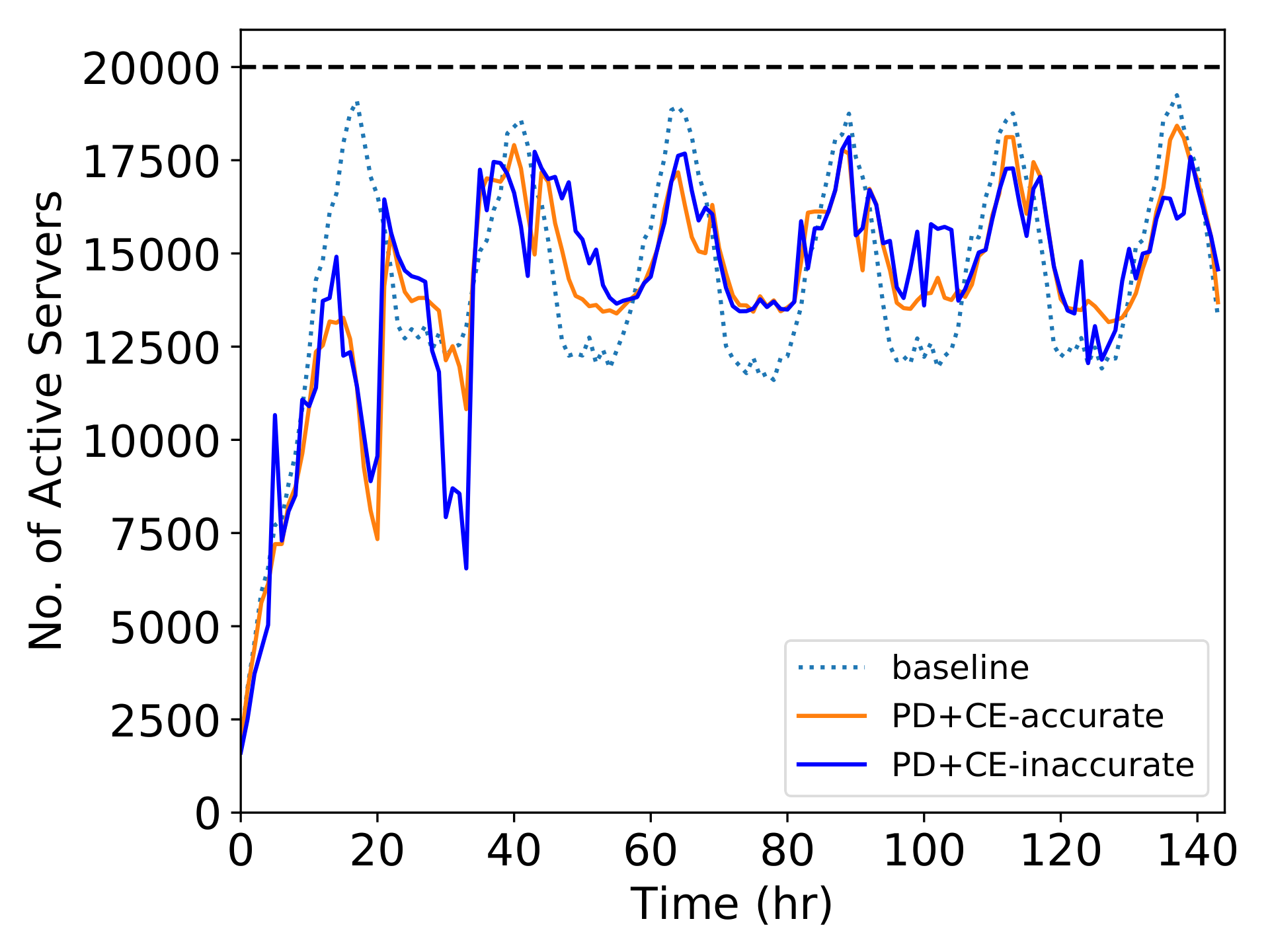}
         \caption{Small Load Variance}
     \end{subfigure}
     \hfill
     \begin{subfigure}[b]{0.33\textwidth}
         \centering
         \includegraphics[width=\textwidth]{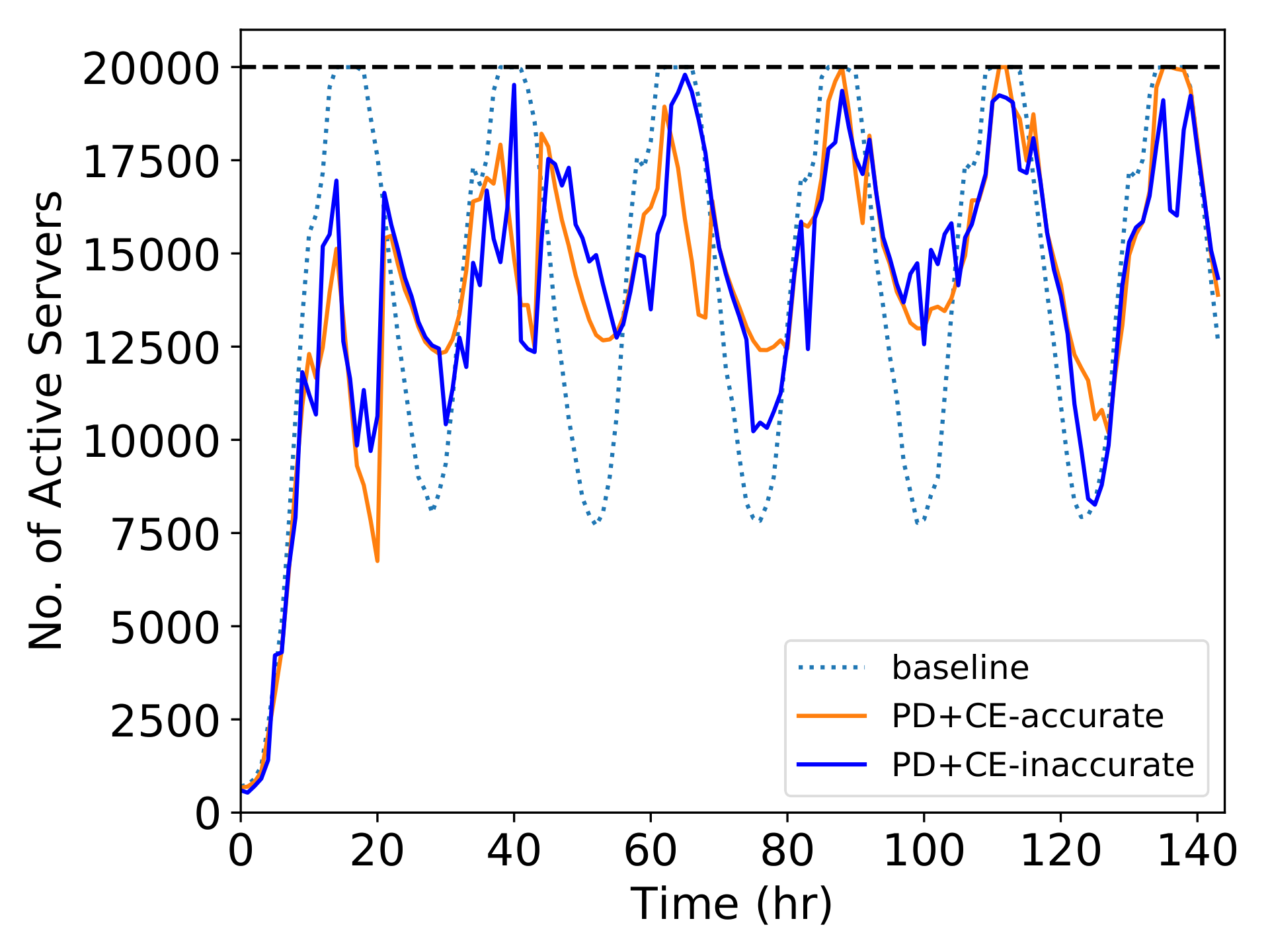}
         \caption{Large Load Variance}
     \end{subfigure}
     \vspace{-15pt}
    \caption{Active server trajectory with accurate vs. inaccurate forecast for carbon emission rates. Baseline denotes the case with $\lambda_{CE} = \lambda_{PD} = 0$. Otherwise, $\lambda_{CE} = 10,  \lambda_{PD} = 100$. }
    \label{fig:inaccurate_carbon}
\end{figure*}

So far, we have assumed that the carbon emission rates available are accurate, even for forecasts. This may rarely be true in reality, where the forecast models lead to certain level of error or uncertainty most of the time. The natural question to address is: does it matter if we use inaccurate carbon emission rates as our model parameters? To answer this question, we generate a series of {\it forecast} carbon emission rates $\bar{cr}$. We assume $\bar{cr}(t) = \xi(t) \cdot cr(t)$, where $\xi(t) \sim \mathcal{N}(1,\,0.11)$ captures the relative prediction error. By construction, $\bar{cr}(t)$ is an unbiased estimator of $cr(t)$ for each $t$. Using the forecast carbon emission rates, we re-run the cases with $[\lambda_{CE},\lambda_{PD}] = [10, 100]$ for each job profile. At each time stage $r$, the DC operator sees the accurate value of $cr(r)$ and forecast values of $cr(t)$ for all $t > r$. 

Figure \ref{fig:inaccurate_carbon} shows the active server trajectory of scheduling with inaccurate forecast, along with the corresponding baselines and scheduling with accurate forecast. The total carbon emission data are shown in table \ref{table:pdce_inaccurate}. We observe that inaccurate forecast does not necessarily lead to further reductions in carbon emissions. This is expected since the forecast is assumed to be an unbiased estimator; there is an equal chance of overestimating versus underestimating future carbon emissions. However, we observe that the active server trajectories with inaccurate forecast can be more volatile compared to those with accurate forecast. This means DCs often need to adjust their schedules to account for sub-optimal schedules decided in previous times due to inaccurate forecast. This demonstrates the value of accurate forecast for DCs even if accurate forecast may not lead to more reduction in carbon mission. On the other hand, this also demonstrates how load-shifting capabilities of DCs enable robust reduction in carbon emission with inaccurate carbon signals.

\begin{table}[h!]
\caption{Carbon emission and volatility comparison for inaccurate forecast cases.}
\label{table:pdce_inaccurate}
\centering
\begin{tabular}{|c|c|c|c|c|c|}
\hline
Job & $\lambda_{CE}$ & $\lambda_{PD}$ & $CO_2$ & $CO_2$ & $\sigma(m(t))$  \\
profile & & & Forecast & ($10^6$ kg) & ($10^3$) \\
\hline
uniform & 0 & 0 & accurate & 6.25 & 1.72 \\
uniform & 10 & 100 & accurate & 6.19 & 2.24 \\
uniform & 10 & 100 & inaccurate & 6.20 & 2.42 \\
\hline
small\_var & 0 & 0 & accurate & 6.29 & 3.18 \\
small\_var & 10 & 100 & accurate & 6.20 & 2.90 \\
small\_var & 10 & 100 & inaccurate & 6.21 & 3.08 \\
\hline
large\_var & 0 & 0 & accurate & 6.32 & 5.11 \\
large\_var & 10 & 100 & accurate & 6.21 & 3.93 \\
large\_var & 10 & 100 & inaccurate & 6.19 & 3.95 \\
\hline
\end{tabular}
\end{table}
\vspace{-15pt}
\subsection{Resource management with varying capacity}
\begin{figure*}[h!]
     \centering
     \begin{subfigure}[b]{0.33\textwidth}
         \centering
         \includegraphics[width=\textwidth]{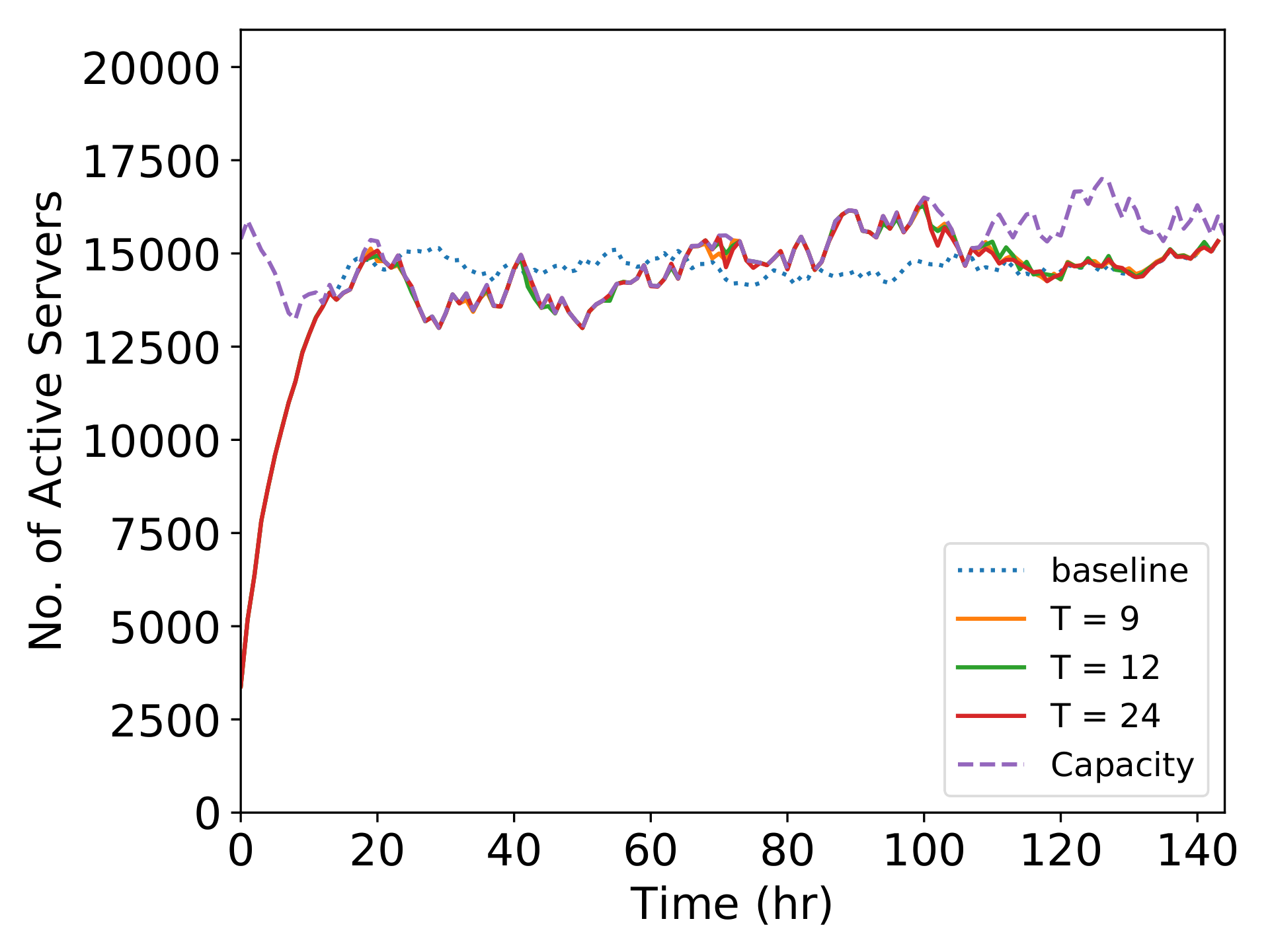}
         \caption{Uniform}
     \end{subfigure}
     \hfill
     \begin{subfigure}[b]{0.33\textwidth}
         \centering
         \includegraphics[width=\textwidth]{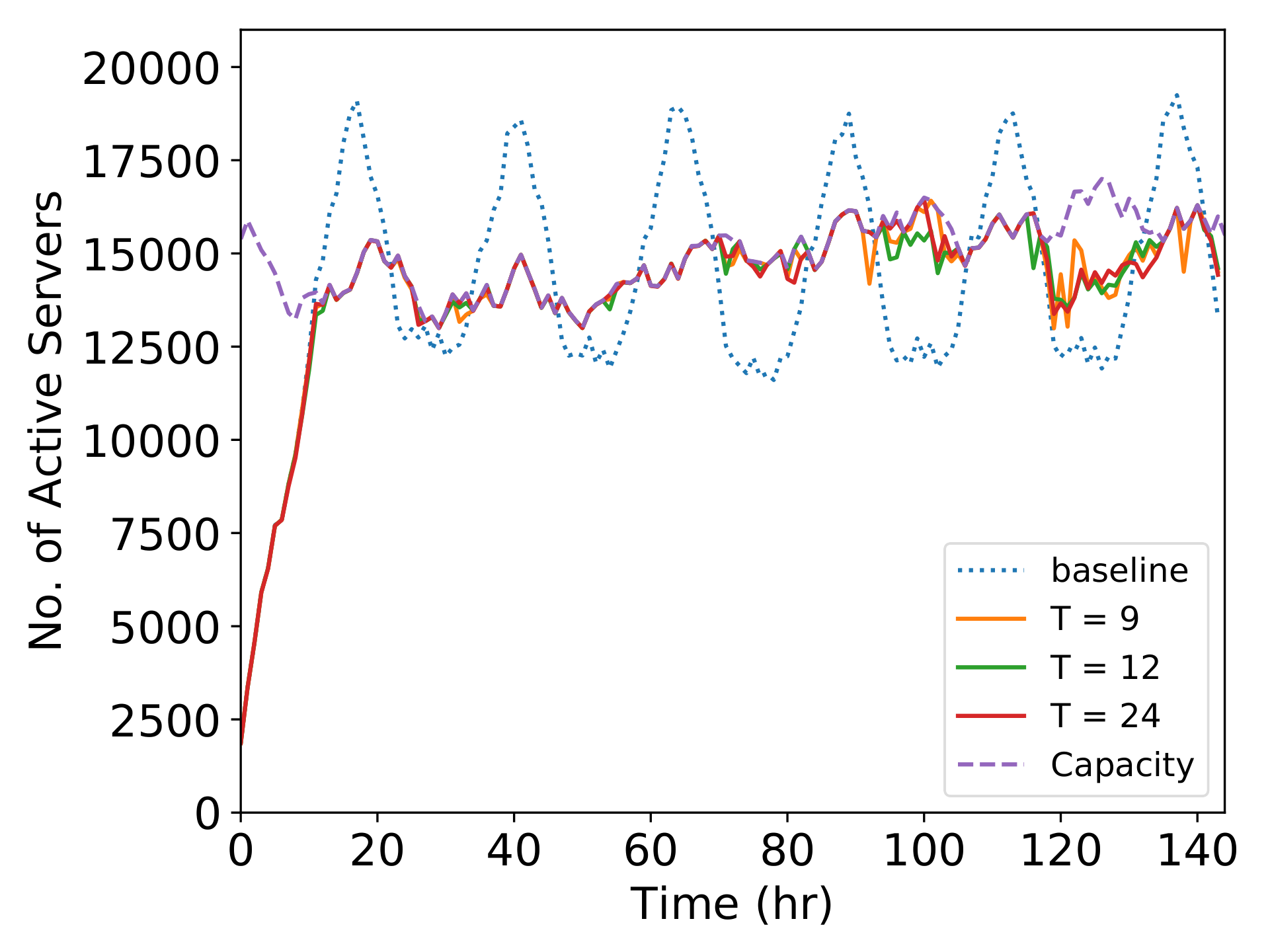}
         \caption{Small Load Variance}
     \end{subfigure}
     \hfill
     \begin{subfigure}[b]{0.33\textwidth}
         \centering
         \includegraphics[width=\textwidth]{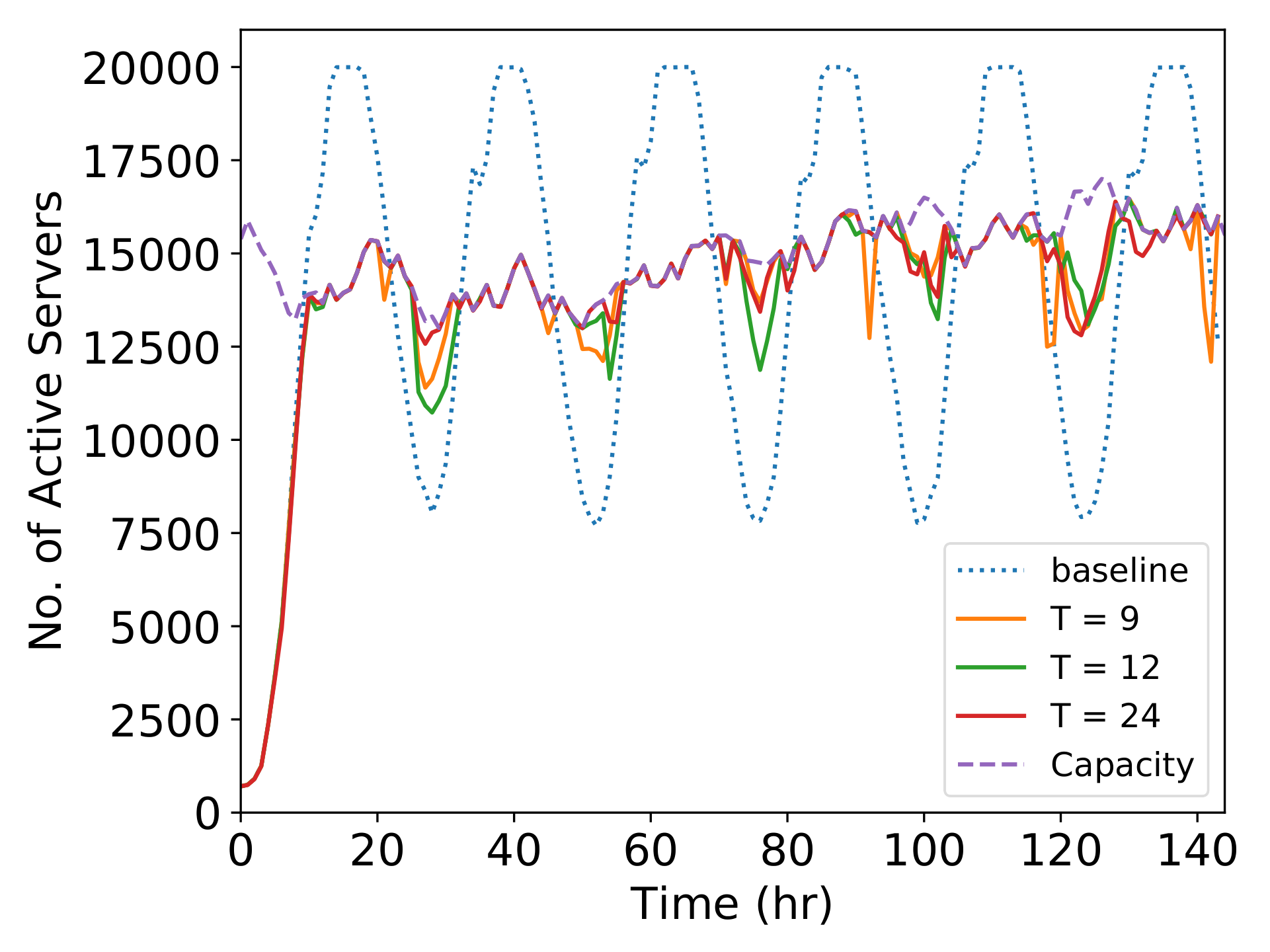}
         \caption{Large Load Variance}
     \end{subfigure}
     \vspace{-15pt}
    \caption{\small Active server trajectory varying computing capacity. $T$ denotes the time horizon values for $T_c$, $T_j$ and $T_h$.}
    \label{fig:varying_cap}
\end{figure*}

\begin{figure*}[h!]
     \centering
     \begin{subfigure}[b]{0.33\textwidth}
         \centering
         \includegraphics[width=\textwidth]{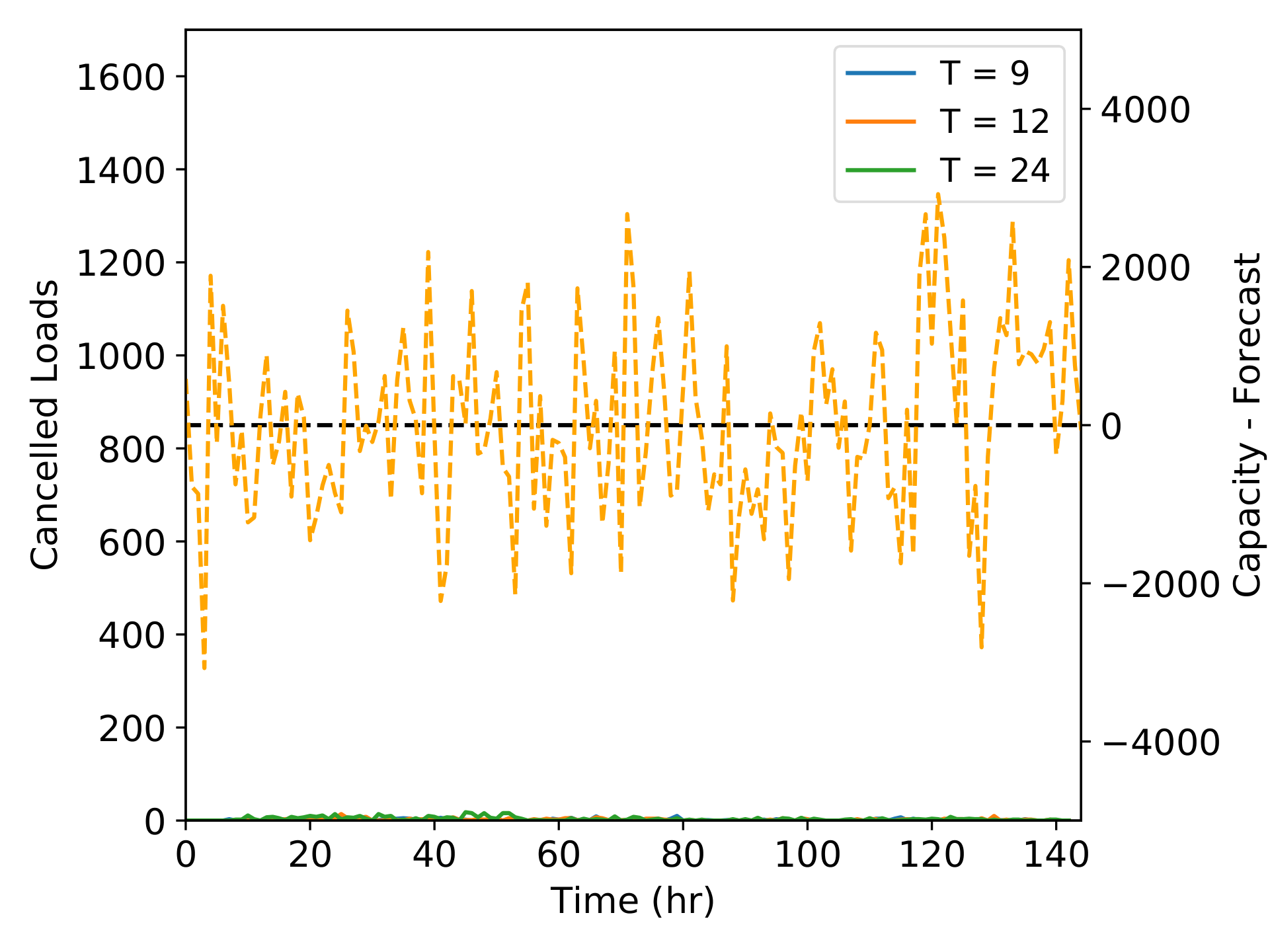}
         \caption{Uniform}
     \end{subfigure}
     \hfill
     \begin{subfigure}[b]{0.33\textwidth}
         \centering
         \includegraphics[width=\textwidth]{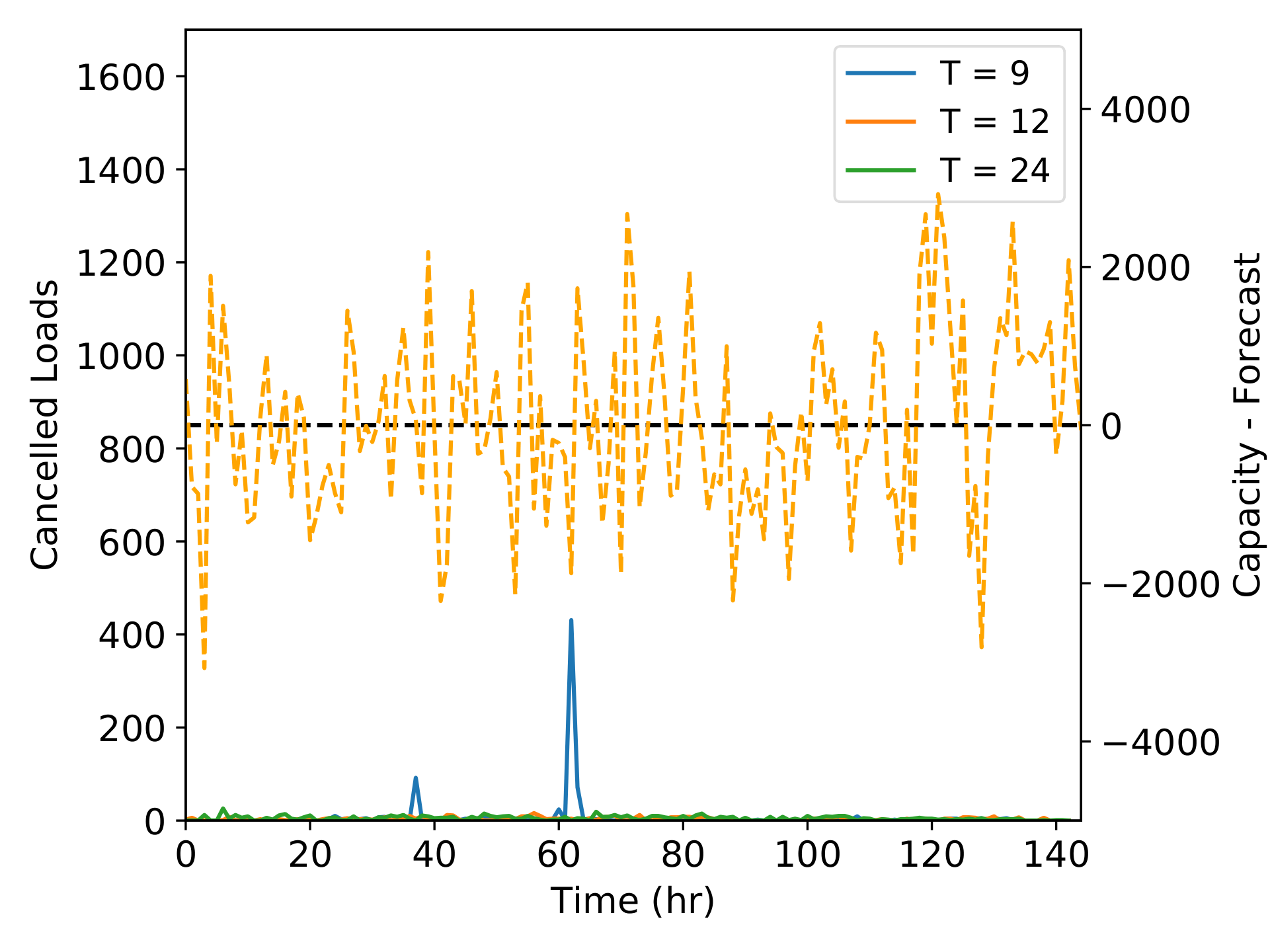}
         \caption{Small Load Variance}
     \end{subfigure}
     \hfill
     \begin{subfigure}[b]{0.33\textwidth}
         \centering
         \includegraphics[width=\textwidth]{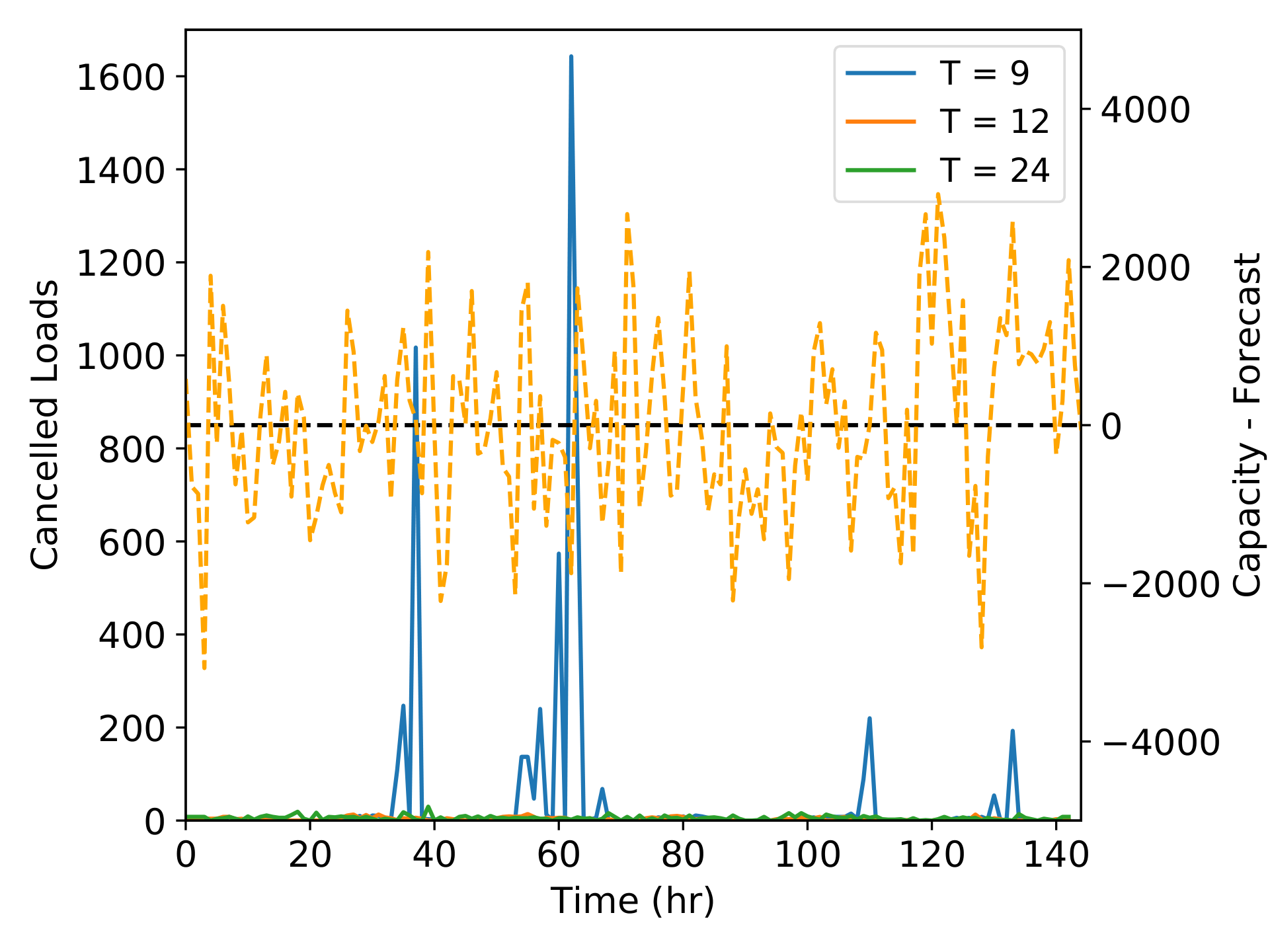}
         \caption{Large Load Variance}
     \end{subfigure}
     \vspace{-15pt}
    \caption{\small Job termination trajectories (solid) and capacity forecast errors (dashed). $T$ denotes time horizon values for $T_c$, $T_j$ and $T_h$.}
    \label{fig:termination}
\end{figure*}

In this subsection, we consider the case where the resource management layer does not necessarily have access to all the computing capacity all the time. Instead, the availability of servers is subject to external conditions such as power dispatch from the grid. This means the operations of DCs are subject to a variable capacity and adds another level of uncertainty to the operations. To simulate this condition, we generate a trajectory $I_t$ for the number of servers available via random walk, and from that, a prediction $\bar{I}(t)$ for the number of servers available $\bar{I}_t = \xi(t) \cdot I_t$ where $\xi(t) \sim \mathcal{N}(1,\,0.07)$. Then we run the cases with $T_h = T_c = T_j = T \in \{9, 12, 24\}$. Figure \ref{fig:termination} shows the number of servers running with jobs that are turned inactive due to job termination at each time. The dashed curve plots the $I_t - \bar{I}_t$, so a negative value means the DC operator overestimates the number of servers available at time $t$. We observe that significant job termination occurs only when $T = 9$, meaning that increasing the decision horizon helps alleviate the need for job termination (and thus becomes more robust to forecast errors). In addition, we observe that job termination usually follows from (possibly consecutive) overestimates for the number of available servers. The active server trajectories in Figure \ref{fig:varying_cap} shows that at the time of job termination, the DCs operator uses all available servers. These are expected since from our formulation, significant job termination should occur only when it is necessary to cut down the number of running servers. In reality, this refers to situations such as power failure. Finally, we observe that as incoming jobs become more volatile, we observe much more job termination. This is expected since more volatile job profiles can see large amount of jobs at certain time intervals, which make it more possible for DCs to see the need for job termination. 
\vspace{-10pt}
\section{Conclusions and Future Work}
In this paper, we propose a receding-horizon, mixed-integer model for DCs that simulates the interaction between the resource management layer and the grid (electricity market) side. The model captures job scheduling constraints to account for the effect of scheduling logics and forecast capability on flexibility provision. The model also incorporates a versatile load aggregation scheme to trade-off model fidelity and computational scalability. The case studies show that carbon-aware hyperscale DCs are able to offer significant flexibility from load shifting in exchange for reduced peak demand charge and carbon emission. In addition, these benefits are robust against carbon signal uncertainty.

As part of future work, we propose to extend our scheduling framework to the setting of geo-distributed DCs, where the DCs can harness spatial load-shifting flexibility too. Along with that, we may assess the effect of various incentives for achieving sustainable operations using our receding-horizon scheduling framework. In this paper we experimented with emission rates, which are straightforward signals and effective from our studies. \blue{However, if we consider overheads of doing load shifting other than job delays (e.g. reduction in reliability), other incentives might be neeeded to overcome these overheads, and our model can be extended to study the effect of these overheads and how markets can remunerate for them. }
\vspace{-10pt}
\section*{Acknowledgments}
The authors acknowledge support from the U.S. National Science Foundation under award 1832208.
\vspace{-10pt}
\bibliographystyle{IEEEtran}
\bibliography{IEEEabrv, refs}

\begin{IEEEbiographynophoto}{Weiqi Zhang}
received the B.S. degree in chemical and biomolecular engineering from University of California, Los Angeles in 2018. He is currently pursuing the Ph.D. degree with the Department of Chemical and Biological Engineering, University of Wisconsin-Madison. His research interests include electricity markets, system flexibility, and optimization.
\end{IEEEbiographynophoto}
\begin{IEEEbiographynophoto}{Line A. Roald}
(Member, IEEE) received the B.Sc.
and M.Sc. degrees in mechanical engineering and
the Ph.D. degree in electrical engineering from ETH
Zurich, Switzerland. She is an Assistant Professor
and Grainger Institute Fellow with the Department
of Electrical and Computer Engineering, University
of Wisconsin–Madison. Her research interests focus
on modeling and optimization of electric grids and
energy systems, with a particular focus managing
uncertainty and risk from renewable energy variability and large-scale outages. She was a recipient of
the National Science Foundation CAREER Award.
\end{IEEEbiographynophoto}
\begin{IEEEbiographynophoto}{Victor M. Zavala} 
received the B.Sc. degree from Universidad Iberoamericana, Mexico City, Mexico, in 2003 and the Ph.D. degree from Carnegie Mellon University in 2008, both in chemical engineering.
He is the Baldovin-DaPra Professor with the Department of Chemical and Biological Engineering, University of Wisconsin- Madison and a Senior Computational Mathematician at Argonne National Laboratory. His research interests are in the areas of energy systems, high-performance computing, stochastic programming, and predictive control.
\end{IEEEbiographynophoto}

\end{document}